\documentclass[12pt]{article}

\usepackage[utf8]{inputenc}
\usepackage{charter}
\usepackage{fullpage}
\usepackage{hyperref}
\usepackage{graphicx}
\usepackage{lipsum}
\usepackage{amssymb}
\usepackage[sort&compress,numbers]{natbib}
\usepackage{hyperref}
\usepackage{verbatim}
\usepackage{natbib}
\usepackage{graphicx}
\hypersetup{hidelinks}
\usepackage{aas_macros}

\usepackage[total={6.5in,9in},top=1in,headsep=0.1in,headheight=1in]{geometry}

\usepackage{authblk}
\usepackage[dvipsnames]{xcolor}
\hypersetup{
    colorlinks=true,
    linkcolor=blue,
    filecolor=magenta,      
    urlcolor=cyan,
    pdftitle={Overleaf Example},
    pdfpagemode=FullScreen,
    }

\usepackage{orcidlink} 

\def\aap{A\&A}

\def\apjs{ApJS}

\newcommand\snowmass{
\begin{center}
  \rule[-0.2in]{\hsize}{0.01in}\\
  \rule{\hsize}{0.01in}\\
  \vskip 0.1in
  Submitted to the Proceedings of the US Community Study\\ 
  on the Future of Particle Physics (Snowmass 2021)\\
  \rule{\hsize}{0.01in}\\
  \rule[+0.2in]{\hsize}{0.01in}\\[-2em]
\end{center}
}

\newcommand{\lcdm}{$\Lambda$CDM}
\newcommand{\wcdm}{$w$CDM}

\usepackage[firstpage=true]{background}
\backgroundsetup{contents={\parbox{6.5in}{\snowmass}}, scale=1,placement=top,opacity=1,color=black,position={3.25in,1.2in}}

\usepackage{fancyhdr}
\fancypagestyle{plain}{%
  \fancyhf{}%
  \fancyhead[C]{}
  \fancyfoot[C]{\thepage}
}

\fancypagestyle{empty}{%
  \fancyhf{}%
  \fancyhead[C]{{\it Snowmass2021 Cosmic Frontier White Paper}}
  \fancyfoot[C]{\thepage}
}
\usepackage[width=0.8\textwidth,font=footnotesize]{caption}

\title{Snowmass2021 Cosmic Frontier CF6 White Paper: Multi-Experiment Probes for Dark Energy -- Transients }
\date{}

\usepackage{authblk}

\author[1]{Alex G.\ Kim\,\orcidlink{0000-0001-6315-8743}}          
\author[2,3]{Antonella Palmese\,\orcidlink{0000-0002-6011-0530}}   
\author[4]{Maria E.\ S.\ Pereira\,\orcidlink{0000-0002-7131-7684}} 
\author[1]{Greg Aldering}
\author[5]{Felipe Andrade-Oliveira}
\author[6]{James Annis}
\author[1]{Stephen Bailey}
\author[7]{Segev BenZvi}
\author[8]{Ulysses Braga-Neto}
\author[9]{Fr\'ed\'eric Courbin} 
\author[5]{Alyssa Garcia}
\author[10]{David Jeffery} 
\author[11]{Gautham Narayan}
\author[1,2]{Saul Perlmutter}
\author[5]{Marcelle Soares-Santos}
\author[12]{Tommaso Treu} 
\author[8]{Lifan Wang} 
\affil[1]{Physics Division, Lawrence Berkeley National Laboratory, Berkeley, 94720 CA}
\affil[2]{Department of Physics, University of California Berkeley, 366 LeConte Hall MC 7300, Berkeley, CA, 94720, USA}
\affil[3]{NASA Einstein Fellow}
\affil[4]{Hamburger Sternwarte, Universit\"{a}t Hamburg, Gojenbergsweg 112, 21029 Hamburg, Germany}
\affil[5]{Department of Physics, University of Michigan, Ann Arbor, MI 48109, USA}
\affil[6]{Fermi National Accelerator Laboratory, P. O. Box 500, Batavia, IL 60510, USA}
\affil[7]{Department of Physics and Astronomy, University of Rochester, Rochester, NY 14627, USA}
\affil[8]{Department of Physics and Astronomy, Texas A\&M University, College Station, TX 77845}
\affil[9]{Institute of Physics, Laboratory of Astrophysics, Ecole Polytechnique 
F\'ed\'erale de Lausanne (EPFL), \protect\\ Observatoire de Sauverny, 1290 Versoix, 
Switzerland}
\affil[10]{Department of Physics and Astronomy, University of Nevada}
\affil[11]{University of Illinois at Urbana-Champaign, Urbana, IL, 61801, USA}
\affil[12]{Department of Physics and Astronomy, University of California, Los Angeles, CA 90095-1547, USA}

\author[13]{\textit{Endorsers:} Odylio D. Aguiar}
\affil[13]{Instituto Nacional de Pesquisas Espaciais, 12227-010, São José dos Campos, São Paulo, Brazil}

\author[14, 15, 16]{Yashar Akrami}
\affil[14]{CERCA/ISO, Department of Physics, Case Western Reserve University, 10900 Euclid Avenue, Cleveland, OH 44106, USA}
\affil[15]{Department of Physics, Imperial College London, Blackett Laboratory, Prince Consort Road, London SW7 2AZ, United Kingdom}
\affil[16]{Laboratoire de Physique de l'\'Ecole Normale Sup\'erieure, ENS, Universit\'e PSL, CNRS, Sorbonne Universit\'e, Universit\'e de Paris, F-75005 Paris, France}

\author[17]{Simon Birrer}
\affil[17]{Kavli Institute for Particle Astrophysics and Cosmology and Department of Physics, Stanford University, Stanford, CA 94305, USA}

\author[18]{Cl\'ecio R. Bom}
\affil[18]{Centro Brasileiro de Pesquisas F\'isicas, Rua Dr. Xavier Sigaud 150, CEP 22290-180, Rio de Janeiro, RJ, Brazil}

\author[6,19]{Elizabeth Buckley-Geer}
\affil[19]{Department of Astronomy and Astrophysics, University of Chicago, Chicago, IL 60637, USA}

\author[20]{Nazim Cabuk}
\affil[20]{Faculty of Engineering, Department of Physics Engineering, Ankara University, Turkey}

\author[21]{Senem Cabuk}
\affil[21]{Faculty of Science, Department of Astronomy and Space Sciences, Ankara University, Turkey}

\author[22]{Laura Cadonati}
\affil[22]{Center for Relativistic Astrophysics and School of Physics, Georgia Institute of Technology, Atlanta, GA 30332, USA}

\author[23]{Mesut Caliskan}
\affil[23]{The William H. Miller III Department of Physics and Astronomy, Johns Hopkins University, Baltimore, MD 21218, USA}

\author[24]{Thomas Y Chen}
\affil[24]{Columbia University, New York, NY 10027, USA}

\author[25]{Tamara M. Davis}
\affil[25]{School of Mathematics and Physics, University of Queensland, Brisbane, QLD 4101, Australia}

\author[26]{Johannes R. Eskilt}
\affil[26]{Institute of Theoretical Astrophysics, University of Oslo, P.O. Box 1029 Blindern, N-0315 Oslo, Norway}

\author[5]{Johnny Esteves}

\author[1]{Simone Ferraro}

\author[27]{Robert Fisher}
\affil[27]{Department of Physics, University of Massachusetts Dartmouth, 285 Old Westport Road, North Dartmouth, MA 02747, USA}

\author[28]{Vadim N. Gamezo}
\affil[28]{Laboratories for Computational Physics and Fluid Dynamics, Naval Research Laboratory, Washington, DC 20375, USA}

\author[29]{Juan Garc\'ia-Bellido}
\affil[29]{Instituto de F\'isica Te\'orica UAM/CSIC, Universidad Aut\'onoma de Madrid, 28049 Madrid, Spain}

\author[30]{Mandeep S. S. Gill }
\affil[30]{SLAC National Accelerator Laboratory, 2575 Sand Hill Road, Menlo Park, CA 94025, USA} 

\author[31]{Kenneth Herner}
\affil[31]{Scientific Computing Division, Fermi National Accelerator Laboratory, Batavia, IL, 60510 USA}

\author[5]{Dragan Huterer}

\author[32]{Mustapha Ishak}
\affil[32]{Physics Department, University of Texas at Dallas, Richardson, TX, 75080} 

\author[33]{Bala Iyer}
\affil[33]{ICTS-TIFR, Bangalore 5600089, India}

\author[34]{Shang-Jie Jin}
\affil[34]{Department of Physics, College of Sciences, Northeastern University, Shenyang 110819, China}

\author[35]{Charles D. Kilpatrick}
\affil[35]{Center for Interdisciplinary Exploration and Research in Astrophysics (CIERA), Northwestern University, Evanston, IL 60208, USA}

\author[36]{L\'{e}on V.E. Koopmans}
\affil[36]{Kapteyn Astronomical Institute, University of Groningen, P. O. Box 800, 9700AV Groningen, the Netherlands}

\author[37]{Macarena Lagos}
\affil[37]{Department of Physics and Astronomy, Columbia University, New York, NY 10027, USA}

\author[38]{Ofer Lahav}
\affil[38]{Department of Physics \& Astronomy, University College London, Gower Street, London, WC1E 6BT, UK}

\author[39]{Mart\'{i}n Makler}
\affil[39,18]{International Center for Advanced Studies, Instituto de Ciencias F\'{i}sicas, Universidad de San Mart\'{i}n, Buenos Aires, Argentina}

\author[40,41]{Peter Nugent}
\affil[40]{Computational Cosmology Center, Lawrence Berkeley National Laboratory, Berkeley, CA, 94720, USA}
\affil[41]{Astronomy Department, University of California, Berkeley, CA, 94720, USA}

\author[13]{Rafael C. Nunes}

\author[42]{Alexei Y. Poludnenko}
\affil[42]{Department of Mechanical Engineering, University of Connecticut, Storrs, CT 06269, USA}

\author[43]{David Radice}
\affil[43]{Institute for Gravitation and the Cosmos, Department of Physics, The Pennsylvania State University, University Park, PA 16802, USA}

\author[4]{Suvrat Rao}

\author[44]{Luidhy Santana-Silva}
\affil[44]{NAT-Universidade Cruzeiro do Sul / Universidade Cidade de S{\~a}o Paulo, Rua Galv{\~a}o Bueno, 868, 01506-000, S{\~a}o Paulo, SP, Brazil}

\author[45]{Misao Sasaki}
\affil[45]{Kavli Institute for the Physics and Mathematics of the Universe, University of Tokyo, Chiba 277-8583, Japan}

\author[43]{Bangalore Sathyaprakash}

\author[46]{Arman Shafieloo}
\affil[46]{Korea Astronomy and Space Science Institute, Daejeon 34055, Korea}

\author[47]{Lijing Shao}
\affil[47]{Kavli Institute for Astronomy and Astrophysics, Peking University, Beijing 100871, China}

\author[5]{Nora Sherman}

\author[48]{Rajeev Singh} 
\affil[48]{Institute of Nuclear Physics Polish Academy of Sciences, PL-31-342 Krak\'ow, Poland}

\author[49]{Joshua Smith}
\affil[49]{The Nicholas and Lee Begovich Center for Gravitational-Wave Physics and Astronomy, California State University, Fullerton, USA}

\author[50]{Aaron Tohuvavohu}
\affil[50]{Department of Astronomy and Astrophysics, University of Toronto, ON, Canada}

\begin{document}

\maketitle

\vspace{3cm}

\abstract{
This invited Snowmass 2021 White Paper highlights the power of joint-analysis of astronomical transients
in advancing HEP Science and presents research activities that can realize the 
opportunities that come  with current and upcoming projects.
Transients of interest include gravitational wave events, neutrino events,
strongly-lensed quasars and supernovae, and Type~Ia supernovae specifically. 
These transients can serve as probes of cosmological distances in the Universe and as cosmic
laboratories of extreme strong-gravity, high-energy physics.
Joint analysis refers to work that
requires significant coordination from multiple experiments or facilities so encompasses Multi-Messenger Astronomy and
optical transient discovery and distributed
follow-up programs.
}

\newpage

\tableofcontents

\newpage

\section{Executive Summary}

A broad range of transient science
requires diverse data sets that can
only be
obtained from multiple experiments/surveys.
Optical telescopes are needed to associate
transient counterparts of gravitational wave standard-siren
discoveries made
by LIGO \& Virgo to measure the Hubble constant $H_0$,
and the neutrino masses from the IceCube experiment. 
Spectroscopic, near-infrared, and enhanced
temporal sampling are needed to get
precise and accurate distances of Type~Ia
supernovae discovered by the Rubin Observatory
to measure the properties of the Dark Energy responsible for the accelerating expansion
of the Universe. 
Similarly, high spatial-resolution
and enhanced
temporal sampling are needed to get precise time delays and modeling
 of strongly-lensed systems discovered by the Rubin Observatory
to measure the Hubble constant. 
The peculiar velocities derived from the
distances of
standard sirens and supernovae can be
compared with the density perturbations
within the same volume
as measured by DESI 
to measure the strength and
length-scale of gravity.

Multi-experiment time-domain science engenders
new considerations that
do not arise in a self-contained experiment.
One facet is experimental design.
Designs can be optimized
for a joint, rather than stand-alone project.
A joint analysis of low-level data products (e.g.\ pixels)
can preserve significantly more
information than the combination
of lossy final data products.
Infrastructure for
real-time inter-experiment
communication can be needed.

The community must now address inconsistencies between
different experiments and/or cosmological probes. The Hubble constant as measured from the
cosmic microwave background, 
baryon acoustic oscillations,
and Type~Ia supernovae are in
tension.  The solution may lie in
new fundamental physics, unaccounted astrophysics, or
experimental systematic errors.

New support 
is needed to enable this time-domain science.
Multi-experiment analysis
can require human and computing resources
beyond the sum allocated to the
individual experiments.
Simulations that
account for
different probes, and
not just  a single experiment,
are needed to interpret
the multi-experiment data self-consistently.
New experiments
must be developed and supported
when existing experiments are insufficient.

\newpage

\section{Dark Energy in the New Era of Multi-Messenger Transients with Gravitational Waves}

\textbf{Coordinator(s):} Antonella Palmese, Maria E. S. Pereira 
\\
\\
\textbf{Contributor(s):} Felipe Andrade-Oliveira, James Annis, Alyssa Garcia, Antonella Palmese, Maria E. S. Pereira, Marcelle Soares-Santos 

\subsection{Dark Energy in a nutshell}
\label{desec1}

The accelerated expansion of the present-day Universe \cite{1998AJ....116.1009R,1999ApJ...517..565P} is  one of the most challenging puzzles in contemporary physics \cite{2008ARA&A..46..385F}. The most straightforward theoretical explanations require a new -- beyond the standard model -- component of the Universe with physical properties that lead to repulsive gravity. This new component is named dark energy and accounts for about $70\%$ of the mass-energy in the Universe. As the high-energy physics (HEP) community top aspiration is to understand how the Universe works at its most fundamental level, dark energy is a high-priority subject of our research program. 

Many theoretical models have been proposed as attempts to provide a compelling explanation for the nature and magnitude of dark energy. For example, a spatially homogeneous, slow-rolling scalar field can provide a negative pressure that drives the cosmic acceleration. This light scalar field could provide a possible mechanism for explaining dark energy. For a comprehensive overview of these models, we refer the reader to the Snowmass White Paper ``Cosmology Intertwined'' \cite{2022arXiv220306142A}.
In this section, we will be considering three of the most popular dark energy parameterizations within a cosmological model, which can be summarized as: 

\begin{itemize}
    \item \lcdm: In this model, dark energy is a cosmological constant ($\Lambda$), and it provides a good fit for most of the observations currently available. Cold dark matter (CDM) is the other  main ingredient of this model. If dark energy is indeed $\Lambda$, its equation of state (EoS) parameter is $w = -1$ at all times. Although  $\Lambda$CDM provides a satisfactory description of the Universe's behavior, e.g. the emergence of late-time cosmic acceleration and structure formation, this model suffers from what appears to be a fatal flaw: assuming that $\Lambda$ corresponds to the energy density of empty space, it can be shown through quantum field theory (QFT) calculations that its value should be many orders of magnitude away from the observed one. This problem motivates the pursuit of alternative models such as those discussed below.

    \item \wcdm: In this more generic model, we let the dark energy EoS parameter $w$ be free to vary. In the \wcdm{} model, the simplest parameterization corresponds to the case of $w$ assuming a constant value over time. 
    
    \item CPL: This model describes a dynamical dark energy. It represents a more complex, but likely more physical, scenario in which the EoS varies with time. This evolution of $w$, in this model, is often described as a function of the scale factor $a$ by the so-called Chevallier–Polarski–Linder (CPL) parameterization: $w(a) = w_0 + w_a (1-a)$, where $w_0$ and $w_a$ are free parameters and the scale factor is defined as $a = (1+z)^{-1}$. Sometimes it is also called $w_0w_a$CDM model. One advantage of this parameterization is that it can provide a good description for a scalar field EoS; that is, in the case dark energy is a dynamical fluid with an EoS that varies with time. 
\end{itemize}

\subsection{The report of the Dark Energy Task Force revisited}

We can measure the impact of dark energy on cosmological observations through its influence on the large-scale structure and dynamics of the Universe. 
Based on this fact, a series of cosmic surveys has been pursued by the HEP community with the goal of achieving a better understanding of dark energy. This series of experiments was outlined in the report of the Dark Energy Task Force (DETF), published in 2006 \cite{2006astro.ph..9591A}.

The DETF was established in the early 2000's by the Astronomy and
Astrophysics Advisory Committee (AAAC) and the High Energy Physics Advisory Panel (HEPAP) as a joint sub-committee to advise the Department of Energy (DoE), the National Aeronautics and Space Administration (NASA), and the National Science Foundation (NSF) on a experimental program for
dark energy research \cite{2006astro.ph..9591A}. The DETF report served as a guide to develop the landscape of dark energy experiments throughout the last two decades including the DES, DESI, JWST, and LSST.
Here, we revisit the major findings of the DETF and present an updated view of the dark energy experimental program for the next two decades. 

Cosmic surveys for dark energy rely on two main classes of cosmological observables. In the first class, known as geometric methods, we study the effect of dark energy on the expansion rate of the Universe as a whole. It can be probed through the distance-redshift relation using standard candles such as type Ia supernovae, standard rulers such as the baryonic acoustic oscillations in the large-scale distribution of galaxies, and more recently, using gravitational wave (GW) as standard sirens. 
The second class of observations relies on measuring the impact of dark energy on the growth rate of large-scale structures. For example, we have measurements of the weak lensing effect on the large-scale structures (e.g. cosmic shear), redshift-space distortions in the distribution of galaxies, and measurements of the abundance of galaxy clusters. A major finding of the DETF is the fact that we can significantly improve constraints on the dark energy EoS by combining these two classes of methods. 

Following the DETF report, the dark energy experimental landscape has been designed to probe dark energy using a diverse set of instruments (e.g. imaging, spectroscopic) capable of probing multiple observables. The DETF report organized these experiments in four categories (Stages),  corresponding to increasing levels of constraining power on the EoS of dark energy.  Here, inspired by the DETF report, we expand the definition of the dark energy experiments' stages into the future: 

\begin{itemize}
    \item Stage I: represents what was known at the time of the DETF report.  
    \item Stage II: represents the then anticipated state of knowledge upon completion of projects that were in progress at the time.  
    \item Stage III: comprises short-term, low/medium-cost projects proposed back then.
    \begin{itemize} 
        \item At the time of this writing, most Stage III experiments are completed or close to completion. 
        \item Examples include: DES, SDSS-III, SHOES, Planck. 
    \end{itemize}
    \item Stage IV: 
    comprises long-term, medium/high-cost projects proposed back then or soon thereafter.
        \begin{itemize} 
        \item At the time of this writing, Stage IV experiments are taking data or in construction. 
        \item For the purpose of this study, we define Stage IV as the anticipated state of knowledge upon completion of these ongoing projects. 
        \item Examples include: DESI, JWST, LSST.
        \item Note that Stage IV was the last stage described in the DETF report. The next stages in this list are newly proposed stages defined in this White Paper for the purpose of guiding the next generation of dark energy experiments.
    \end{itemize}
    \item Stage V: 
    near-term, low/medium-cost future projects.
    \begin{itemize} 
        \item This is the first new Stage in the series. It represents beyond Stage IV experiments of medium scale, which are proposed to be completed in a 5-10 years timescale.    
        \item Examples: MegaMapper, MSE, DESI-2, CMB-S4, LIGO Voyager. 
    \end{itemize}
    \item Stage VI: 
    long-term, medium/high-cost future projects.  

    \begin{itemize} 
        \item This is the second additional Stage in the series. It represents future experiments of larger scale and longer time frame. 
        \item Examples: Cosmic Explorer, Einstein Telescope. 
    \end{itemize}
\end{itemize}

\subsection{New challenges and goals for dark energy research in the next two decades}

The figure of merit used by the DETF to compare different experiments was the inverse of the area constrained in the $w_0 \times w_a$ parameter space. This figure of merit is no longer sufficient to capture the potential of future experiements to achieve our current dark energy science goal. Beyond measuring the EoS with increased precision, we  aim at {\it distinguisihing} between dark energy models. For example, we want to be able to confidently say whether or not the universe is currently dominated by a dynamical dark energy fluid. In other words, we want to be able to reliably test/falsify the $\Lambda$CDM hypothesis. 

A path towards this goal is to explore the observed tensions between results arising from different observables and experiments. Measurements of the rate at which the Universe is expanding today (i.e. the Hubble constant, $H_0$) derived from the cosmic microwave background (CMB) and late-time observations exhibit a significant tension at a level of $\sim 4 - 6 \sigma$ \cite{2019NatAs...3..891V}. Moreover,  measurements of the matter fluctuation amplitude $\sigma_8$ and the matter density $\Omega_m$ from galaxy cluster samples and lensing present systematically lower values in comparison with the results from CMB \cite{2019arXiv190105289D}. 

With the experiments and precision that we have today, we cannot distinguish whether these observed tensions are due to unaccounted-for systematic uncertainties or if they are evidence that \lcdm{} is incorrect or incomplete. In particular, performing measurements of the late time Universe with highly non-linear physics, combined with the difficulty of adequately accounting for systematic errors, are a difficult task even in the timeline of the Stage IV experiments \cite{2006astro.ph..9591A}. Therefore, it is essential to look for new independent observables and to perform robust joint analyses \cite{2022arXiv220306795B}. 

One goal for future experiments is to go beyond achieving per cent-level precision to identify statistically significant tensions between the multiple observables for dark energy in order to test and exclude dark energy models if they are not realized in nature. In order to quantify progress towards this goal, we define a new figure of merit to compare different proposed experiments for Stages V and VI dark energy science. This new figure of merit is defined as the \textit{tension significance} that each experiment would achieve, if instead of \lcdm{}, dark energy is better described by a time-evolving equation of state. 

In the next sections, we present the potential of transients, particularly gravitational wave standard sirens, to achieve this goal in the next decade.       
\subsection{Dark energy science with gravitational wave standard sirens}

Compact object binary mergers are promising novel probes of cosmology. Often referred to as standard sirens, such events can be used as multimessenger probes. Gravitational waves emitted during the inspiral of compact binary mergers are used to determine their absolute distances, while their redshifts can be determined via traditional astronomical observations, if an electromagnetic counterpart (and/or its host galaxy) is found. Thus, these standard sirens are used as distance indicators, analogous to type-Ia supernovae. 

The primary motivation to pursue standard sirens for cosmology is that, contrary to supernovae, their distances are determined from first principles. By removing the need for multiple astrophysical calibration steps, we eliminate some challenging systematic uncertainties. As we mentioned, out of the several key parameters of modern cosmology, distance indicators are particularly sensitive to the expansion history of the Universe as a function of redshift,  the Hubble parameter $H(z)$. Nearby sources in particular are used to measure the current rate of expansion, $H_0$. A breakthrough in understanding the physics of dark energy is a core goal of our community to understand the accelerated expansion of the cosmos and the absolute value of today's expansion rate is important -- per cent-level precision measurements of $H_0$ are required to cease being a limiting factor on dark energy model limits. 

Compared to the well-established cosmology probes such as supernovae, the emerging field of multimessenger cosmology with gravitational waves is advancing in leaps. Since the first observation \cite{Abbott16} of a compact binary system merger by the Laser Interferometer Gravitational-wave Observatory (LIGO) and Virgo Collaboration (LVC), less than five years ago, over 90 merger events and candidates have been reported \cite{gwtc3}, one has had its electromagnetic counterpart identified \cite{2017PhRvL.119p1101A, Abbott17a, Soares17}, and a handful of events without counterpart have been used for measurements of $H_0$ with an uncertainty of $\sim 14-20\%$ \cite{AbbottH0, 2019ApJ...876L...7S, 2020ApJ...900L..33P, 2021arXiv211103604T,palmese_StS_DESI}. As the GW detector network continues to be upgraded, the sample sizes are expected to grow exponentially in the coming decade. Therefore, the ambitious goal of per cent-level $H_0$ measurements from standard sirens is a realistic possibility. A precision measurement of $H_0$ from nearby GW sources, out to the distances expected for binary neutron star mergers from current generation GW detectors, is also a powerful mean to break degeneracies between cosmological parameters from future CMB experiments (in particular the geometrical degeneracy) or from BAO measurements from e.g. DESI \cite{divalentino2018}. Besides that, detection of gravitational waves events at higher redshifts ($z \sim 0.7$), which is also expected for binary neutron stars with next generation (XG)\footnote{After the Snowmass Community Summer Study Workshop, it was proposed to adopt the initials XG for the next generation of GW detectors instead of 3G.} GW detectors \cite{2022arXiv220308228B}, might allow us to probe other cosmological parameters beyond $H_0$, e.g. the EoS of dark energy. 

Beyond measuring these cosmological parameters with high precision, we propose that an entirely new program be designed to build a large sample of standard sirens with identified electromagnetic counterparts in support of an analysis capable of distinguishing between \lcdm{} and the different dark energy models with high significance. This program will rely on a high-efficiency search and discovery campaign using ground-based telescopes in tandem with the next generation of gravitational waves detectors.  A pioneering version of this program is currently being pursued by the Dark Energy Survey collaboration (the DESGW program \cite{2020A&C....3300425H}) using DECam for imaging and other telescopes for spectroscopic follow-up to confirm candidates.  Here, we do not describe in detail the scope of future observations, but we anticipate that a new telescope/instrument systems will be required to perform rapid follow-up observations. Lessons learned from DESGW will be useful in designing and optimizing such a future observational program.

\subsection{A precursor follow-up instrument for the XG era}
\label{section:decam}

While we anticipate the need for an instrument dedicated for GW transients searches in the 2030s to fully establish a dark energy program with gravitational waves, we also believe that this program will benefit from a near-term -- from now to 2030s -- dedicated optical instrument for follow-up search. As an example, one of the most efficient instruments used for this purpose today is the DOE/DECam at the Blanco Observatory in Chile. DECam's $\sim 3$ sq. deg. field of view allows us to quickly cover the GW probability sky area in $griz$ or $Y$ filters out to the limits necessary to detect an optical counterpart at the distances expected in future GW observing runs. DECam has been used for target of opportunity (ToO) observations by several groups. In particular, the DESGW group has demonstrated the potential of DECam in following up GW events \cite{2020ApJ...903...75G, 2020ApJ...901...83M, 2022ApJ...925...44S, 2022ApJ...929..115T,Bom2022}. Additionally, LSST will be online in the near-term timeline. Although it has a transient program planned, LSST will not provide the most effective set up for \textit{fast} follow up observations. While LSST is not a prime instrument for GW follow ups, its data will be essential for finding GW transients. Then, we argue that coordination between dedicated DECam searches and LSST will enhance our ability to find many GW transients. 

In summary, an existing 3-4m optical instrument could be leveraged up as a precursor instrument for dark energy science through the search and discovery of GW and other transients --- see \autoref{tdm}, for possible synergies with the Time Delay Machine, that could establish them as the two leading new techniques for Hubble constant's measurements in the XG era.

\subsection{Experiments and stages}
\label{section:stages}

\begin{figure}[ht]
\centering
\includegraphics[width=\textwidth]{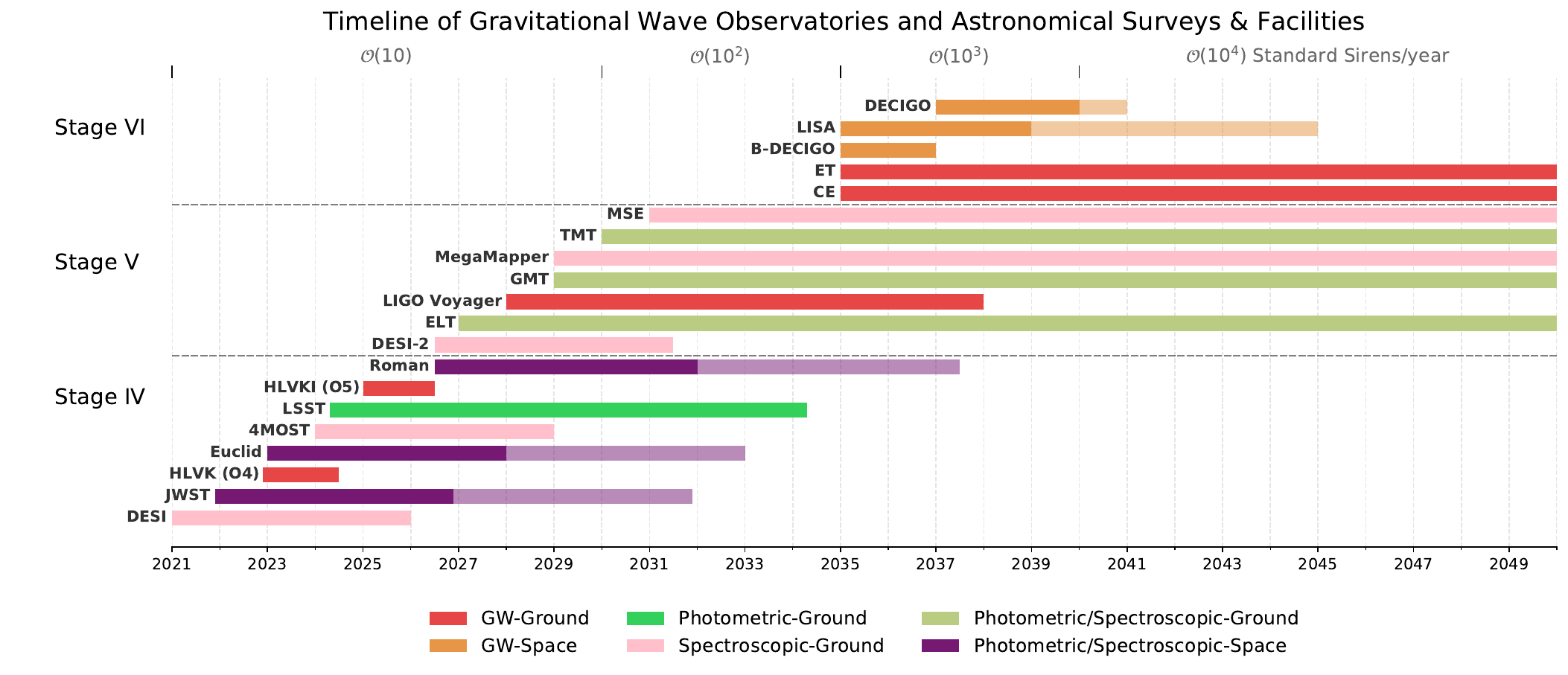}
\caption{Timeline of current and planned gravitational wave detectors and astrophysical surveys/facilities. The horizontal dotted lines indicates the dark energy experiment stages outlined in this White Paper. The darker colors represent the nominal lifetime of the facility, and lighter colors are possible extensions. The numbers in the top are the expected number of standard sirens that will be detected per year based on projections from \cite{borhanian2022listening}. We estimate that standard siren experiments for dark energy must yield samples corresponding to 5-10\% of the total number of events detected by the GW observatories in order to achieve the Stage V and Stage VI goals.} 
\label{fig:timeline}
\end{figure}

To demonstrate the constraining power we may achieve in the future with a DESGW-like program operating in coordination with future GW observatories, we have examined the landscape of currently proposed facilities using our definitions of post-DETF dark energy science stages. An estimated timeline for these stages and experiments is presented in \autoref{fig:timeline}: 

\begin{itemize}
    \item In Stage IV, the LIGO detectors as well as Virgo and KAGRA will be online and will undergo upgrades in order to achieve the proposed LIGO (HLI+), Virgo+ (V+), and KAGRA+ (K+) sensitivies. During this time, telescopes such as LSST \cite{lsst2012large} and DESI will also be operating. 
    \item Next, Stage V will include the HLIVK+ network with the improved LIGO Voyager detector. During this time, we anticipate telescopes such as MegaMapper, MSE and DESI-2 to come online. During this era, GW detectors could be pushing the redshift horizon beyond z $\sim$ 1 for binary mergers. 
    \item Finally in Stage VI, the addition of Cosmic Explorer and the Einstein telescope will bring the ground based detector network sensitivity to $z > 4$. Note that during this era, there are no anticipated telescopes that are planned to be powerful enough for optical follow up of these events. 
\end{itemize}

For the purpose of a dedicated standard siren dark energy program, we need to efficiently detect electromagnetic counterparts (a task requiring  deep and wide imaging as well as spectroscopy). Remarkably, the bottleneck to fully exploiting the potential of standard sirens for dark energy research is in the traditional cosmic survey arena instead of the GW observatories. 
Already in Stage V, it will be challenging to meet those requirements with the proposed experiments. In Stage VI there are no telescopes currently planned for this purpose. In this paper, we show how powerful our dark energy results would be if we can obtain electromagnetic counterparts for about 5-10\% of the events that are expected to be seen by the currently proposed GW observatories. 

\subsection{Forecasts}

In this section, we quantify how powerful standard sirens are as a tool to distinguish between \lcdm\ and two other dark energy models. We explore the induced bias caused by some wrong assumption of the cosmological model and the sensitivity of this cosmological probe to the cosmic expansion rate ($h$, which is $H_0$ in units of 100 km/s/Mpc), the energy density of dark matter ($\Omega_m$) and the dark energy EoS parameters ($w_0$ and $w_a$). 
Note that, until now, standard siren analyses have only been able to measure $h$, fixing the other parameters to the \lcdm\ values, or letting them vary (but without being able to obtain useful constraints). In the future, however, we expect standard sirens to be detected out to greater redshifts, which  will drastically improve their constraining power beyond $h$, provided that successful campaigns for electromagnetic counterpart detection are also pursued.

For this forecast analysis, we simulated three data vectors based on different model universes: \lcdm, 
\wcdm, and CPL. We used the current best-fit values of \lcdm, and we set the EoS parameters of the other two models to be no more than one standard deviation away from the current state-of-the-art constraints \cite{2020A&A...641A...6P,2016A&A...594A..14P}. The only change was in the value of the EoS parameters, all other relevant cosmological parameters are set to the best-fit values of the Planck 2018 results. The chosen parameters are shown in \autoref{tableW}. 

\begin{table}[ht]
\begin{center}
\begin{tabular}{cccc}
\hline
               &$\Lambda$CDM & $w$CDM & CPL \\ \hline
$h$            &     0.67    &    0.67     & 0.67     \\
$\Omega_{\rm m}$ &     0.265  &   0.265     & 0.265      \\
$w_0$     &     -1      &  -0.92     &  -0.877      \\
$w_a$     &      0.0      &      0.0     & 0.12      \\
$\Omega_k$     &      0.0      &      0.0     & 0.0      \\
\end{tabular}
\caption{Parameter choices for GW datavector simulations. We chose values for equation of state parameters, $w_0$ or ($w_0, w_a$), inside the $1\sigma$ confidence level of the current best constraints. All other relevant parameters are set to the Planck 2018 $\Lambda$CDM nominal values \cite{2020A&A...641A...6P}.}
\label{tableW}
\end{center}
\end{table}

The basic idea of this forecast analysis is that, if we assume \lcdm{} for all three universes and attempt to fit the simulated data, we will find $h$ and $\Omega_m$ values that are in significant disagreement with the input. This is because the effect of a dynamic equation of state can be mimicked by changing the values of these two parameters. After defining the key parameters of the observable data set, the forecast is done by determining the significance of the bias caused by the wrong assumption of the cosmological model. If our goal is to falsify \lcdm, we should aim for at least a $5\sigma$ level tension. Ideally, the combination of GW with other independent probes (and therefore, distinct bias) will induce considerable internal tensions in the model. But this approach is not explored in this work.  

The observable data set parameters for Stage IV, V, and VI experiments are defined in \autoref{tableStages}. The key parameters are the number of events and the GW's redshift range. Here we assume the values given by the proposed GW experiments and assume that our search and discovery program for electromagnetic counterparts will be successful in amassing a sample of about 5-10\% of the best GW events. Another important factor in determining the constraining power is, of course, the distance uncertainty. Here we estimate an overal 5\% uncertainty on the distances, as estimated by the GW teams for their top events. 

\begin{table}[ht]
\begin{center}
\begin{tabular}{cccc}
\hline
Stage & GW network type & $z$ range & No.\ of events  \\ \hline
IV    & HLVKI  & 0.5 & 110\\ 
V     & Voyager & 1.0             & 1300  \\
VI    & Cosmic Explorer & 4.0             & 6000  \\
\end{tabular}
\caption{Summary of GW ground-based observatories that are anticipated at each stage. For each stage we list an approximate redshift range and number of events with an identified electromagnetic counterpart per year, which we used for our simulations. In this estimates, we assume a search and discovery program capable of identifying about 5-10\% of the GW events.}
\label{tableStages}
\end{center}
\end{table}

We produced simulated datavectors drawn from each of the cosmological model ($\Lambda$CDM, $w$CDM or CPL) and statistically consistent with the expected observational setup (e.g., number of detections, maximum redshift) for each of the different experiment stages. We then fit each simulated datavector assuming the $\Lambda$CDM model. In this way, we can estimate the disagreement between the input parameters (truth) and the fitted parameters \cite{Shafieloo2018}.

We consider a uniform volume rate of detection, and the maximum redshift of detection as $z=0.5$, $z=1.5$, $z=4.0$ for the stages IV, V, and VI, respectively \cite{borhanian2022listening}. We assume a typical value to the signal-to-noise ratio of SNR=20, implying $\sigma_{d_L}/{d_L}\sim5\%$ and spectroscopic redshift determination of the host galaxy. For this exploratory discussion, we are considering only detection of optimally oriented events, i.e. face-on and overhead with respect to the detectors.  

\begin{figure}[ht]
\centering
\includegraphics[width=0.44\linewidth]{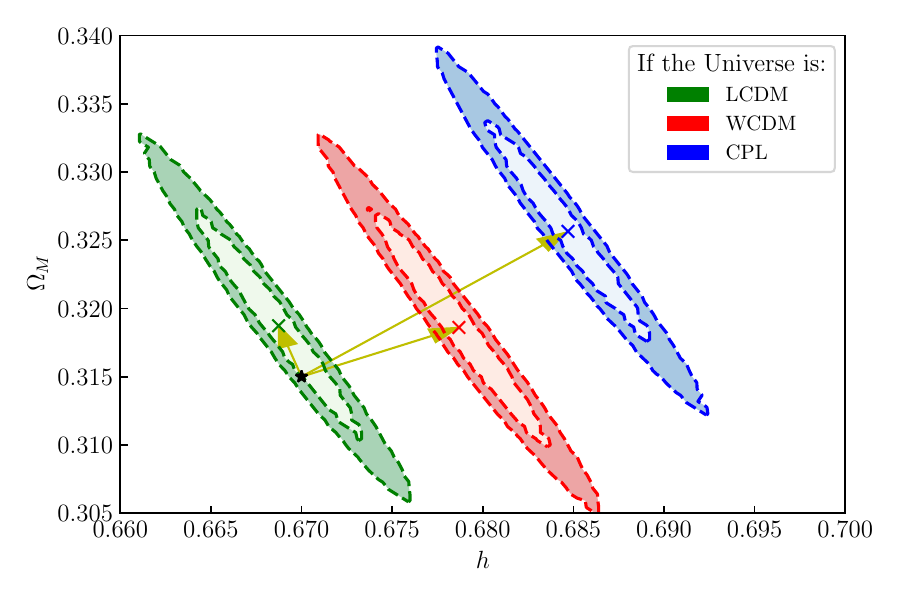}
\includegraphics[width=0.47\linewidth]{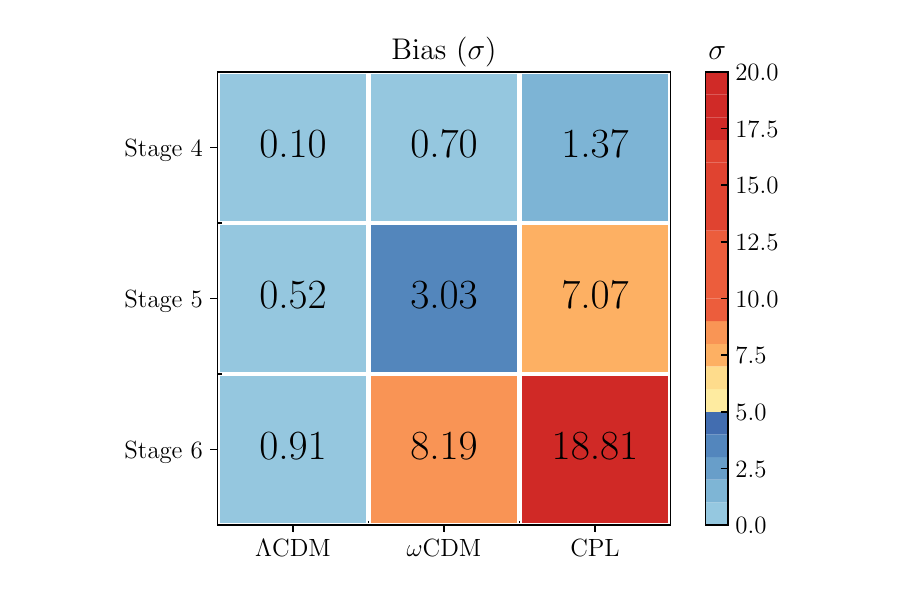}
\caption{\label{fig:tensions}
Left: Parameter constraints using a single realization of a Stage VI-like simulated data. This result illustrates the magnitude of the tension expected in $h-\Omega_m$ space due to the assumption of $\Lambda$CDM as cosmological model. We show the resulting constraints when fitting simulated datavectors with distinct dark energy models, namely, $w$CDM (red), CPL (blue), and the $\Lambda$CDM (green). We also show the truth input parameters for reference (black star). For each of the simulations, we have ($h$, $\Omega_{M}$)= (0.67, 0.315) and the shift is caused by the wrong model assumption.
Right: Assuming $\Lambda$CDM as fitting model, we show the parameter bias in confidence level (c.l.) units when the realization of simulated datavector is based in different cosmological models. For example, for a Stage V experiment, if the true dark energy model in the Universe is CPL, we would exclude \lcdm\ at the $7\sigma$ level. For \wcdm\, the situation is more challenging, and we would need a Stage VI experiment to achieve the same level of confidence in excluding \lcdm.}
\end{figure}

A summary of our results is found in \autoref{fig:tensions}. On the left panel, using a single realization of the simulated data, we show how the different dark energy models affect the result in the $h-\Omega_m$ plane for the Stage VI experiments. Note that there is a very prominent shift between the contours, showing a significant bias in the constrained parameters. We estimated the significance of this bias and repeated this exercise for Stages V and VI. The results are shown on  the right panel of \autoref{fig:tensions}.

 In summary, our results show that if the EoS of dark energy for the observed Universe has a deviation from $w=-1$ in $1\sigma$ c.l., at the end of the Stage V, we may observe a bias with a significance of up to $7\sigma$ in the $h-\Omega_{m}$ plane, which will induce a high level of tension in the combined analysis of GW with other independent probes (e.g. large-scale structure, BAO). For the observational scenario considered in the Stage VI, the deviation in the $(h, \Omega_{m})$ from its true value, may reach $18 \sigma$ of c.l. for the most divergent scenario investigated here. This result shows that standard sirens will be a truly powerful tool for dark energy science in the next two decades, if we are capable of observe a significant fraction of them with the available and proposed facilities.

We note that here we used our forecasting tools specifically to quantify the impact of standard sirens as a novel cosmological probe. The rationale and the 
tools used here can be used in principle for comparison with other models and probes as well. 

\subsection{Discussion}

Standard sirens are a new tool for dark energy research. Our community has just started exploring this new tool and its full potential is yet to be achieved. In this paper, we present a study that makes a strong case for establishing a high-efficiency search and discovery program for GW events with electromagnetic counterparts. 

Inspired by the report of the dark energy task force, we define the new upcoming stages for the field of dark energy: Stage IV are the ongoing experiments, Stage V and Stage IV are the next and next-to-next generation of proposed experiments.

The goal of the HEP community for dark energy science has shifted from  measuring the EoS parameters with greater precision to testing the \lcdm\ paradigm with high-precision and high-accuracy experiments capable of significantly distinguishing between \lcdm\ and other dark energy models. In this context, the new metric for comparison between dark energy experiments is the significance of the bias that could be measured between models. 

In this study, we quantify this statement by estimating the statistical significance that could be achieved in excluding \lcdm\ if the Universe is instead dominated by a dynamic form of dark energy (\wcdm\ and CPL). We assume that the observational data will be typical of the proposed next generation of dark energy experiments, and that our future search and discovery programs will be efficient in identifying counterparts for the top 5-10\% of the events. We showed that standard sirens alone would be able to exclude \lcdm\ with significance well above $5\sigma$ c.l. in this scenario. This result bodes well for the future of multi-messenger cosmology with gravitational wave standard sirens. Observations of surprise  counterparts to some fraction of stellar mass binary black holes could also further improve the constraints presented here \cite{Palmese_2021}.

We stress that the use of GW standard sirens is complementary to that of Supernovae as standard candles. First, by providing an absolute measurement of luminosity distance, standard sirens are free of the cosmic distance ladder systematics that are relevant in Supernova cosmology. More in general, because they are sensitive to a very different set of systematics than standard candles, they provide an independent measurement of cosmological parameters that is relevant where tensions between cosmological parameters arise (e.g. in te case of the Hubble constant tension). Secondly, XG gravitational wave detectors will provide exquisite luminosity distance measurements, with a precision down to 1\%, which Supernovae are not expected to reach due to their intrinsic scatter in luminosity. As such, GW standard sirens from the Cosmic Explorer will allow us to build very precise maps of the nearby Universe, and measure the galaxies' peculiar velocity field to derive constraints on the growth of large scale structure and gravity models \cite{palmese20} (see Section 4 for a description on peculiar velocities from Supernovae). Moreover, observations of LISA sources will be able to extend the Hubble diagram beyond what is achievable with Supernovae (e.g. \cite{speri2021}). 

In order to fully exploit the outstanding potential of standard sirens for dark energy research, we need to efficiently detect electromagnetic counterparts (a task requiring  deep and wide imaging as well as spectroscopy). Remarkably, the bottleneck here is in the traditional cosmic survey arena instead of the GW observatories. In particular, one of the limitations to GW counterpart identification has been the lack of rapid classification of transients found through imaging. This will also be a bottleneck for time domain research from LSST more generically, beyond GW follow-up. Multi-object spectroscopy with e.g. DESI and DESI-2 will be able to provide a solution to this problem \cite{palmese_WP}. We therefore recommend that multi-object spectroscopic experiments, together with wide-field imagers such as LSST, in the future, will have a dedicated program for GW follow-up. Observations by these wide-field multi-object spectrographs can further support GW follow-up campaigns by other instruments with the observations of potential host galaxies in the GW localization region. Once the counterpart is identified, additional precision measurements of the transient can enable further characterization of the GW system geometry, which can be used jointly with GW observations to improve on the distance measurements and hence on the cosmological parameters \cite{guidorzi,KNH02020}. Rapid, spectrophotometric observations from a space telescope, such as the one we mention in Section 4, with a dedicated GW follow-up program, would be ideal for this task.

Already in Stage V, it will be challenging to meet those requirements with the proposed telescope facilities. In particular, because they will not provide capabilities for efficient and fast follow-up observations. Therefore, we argue that will be highly beneficial in this stage to re-purpose one of the existing 4m telescopes (e.g. DECam) as a dedicated instrument for the search and discovery of GW and other transients. In Stage VI there are no telescopes currently planned for this purpose. Based on these results, we propose that the community develops a novel standard siren survey program coordinating with the GW observatories to fully incorporate this new observable into our research portfolio for dark energy science.

\section{Multi-Messenger Physics With Neutrinos}

\textbf{Coordinator(s):} Alex G. Kim 
\\
\\
\textbf{Contributor(s):} Segev BenZvi, Alex G. Kim  
\\

Neutrino events detected by DOE projects such as DUNE, nEXO, LZ, Xenon1T, KamLAND, etc., as well as non-DOE projects such as Super-Kamiokande, IceCube and KM3Net, enable a range of fundamental, gravitational, and astrophysical science when associated with optical and gravitational wave counterparts. Potential high-impact measurements include the detection of possible neutrino decay, bounds on keV-mass sterile neutrinos, tests of Lorentz-invariance violation, measurements of the speed of neutrinos, and observations of neutrino interactions in high-energy environments beyond the reach of human-built accelerators \cite{Engel:2022, Ando:2022}.

The optical component of the suite of multi-messenger data consists of
prompt, highly-cadenced, and multi-wavelength observations that generally require triggered targeted follow-up. Obtaining these data thus requires fast access to telescopes and an accompanying infrastructure that facilitates making scheduling decisions derived from information from a diverse set of independent experiments.

The US HEP program supports surveys with large telescope time allocations equipped with wide-field ($\gtrsim 10$ sq.deg.) imaging and spectroscopic cameras that are ideal for multiplexed observation of tens of thousands to millions of potential transient sources and their host galaxies. For example, DECam has performed triggered follow-up observations of IceCube neutrino events and estimated the production of TeV neutrinos in core-collapse supernovae \cite{DES:2019hqa}, while the highly-cadenced ZTF survey has identified a possible associations between neutrinos and tidal disruption events \cite{Stein:2020xhk, Engel:2022}. As another case in point, DESI, through an internal secondary target time allocation, performed triggered observations \cite{2021GCN.30923....1P} of IceCube event 210922A \cite{IceCube:210922A}. In the next decade, the combination of spectroscopic redshifts at $z\lesssim0.3$ from surveys such as DESI with observations of cataclysmic transients from facilities like DECam and the Vera C. Rubin Observatory Legacy Survey of Space and Time will significantly improve constraints on the origins of astrophysical neutrinos \cite{DES:2019hqa}.

The multi-messenger observations of a burst of $\mathcal{O}(10~\mathrm{MeV})$ neutrinos from a nearby core-collapse supernova, as well as associated gravitational waves and electromagnetic radiation, present a once-in-a-lifetime opportunity to study flavor oscillation effects unique to the dense supernova core \cite{Duan:2010bg, Sasaki:2019jny, Berryman:2022} and a range of physics beyond the Standard Model \cite{Fischer:2016cyd, Muller:2019upo, Franarin:2017jnd, Arguelles:2016uwb}. Matched temporal features in the fluxes of supernova neutrinos and gravitational waves from a nearby core-collapse or a distance compact binary merger would probe the speed of gravitational waves \cite{Nishizawa:2014zna}, and could provide a signature of the formation of a black hole \cite{Pagliaroli:2011zz, Nakamura:2016kkl}. Finally, the observation of supernova neutrinos would provide a crucial early warning minutes to days before the detection of optical signals.

To take full advantage of the scientific opportunity of a nearby core-collapse supernova, support is needed for both theoretical modeling and follow-up infrastructure. Non-linear neutrino flavor mixing in extremely dense media is challenging to simulate but crucial to understanding the observed neutrino signal and the effect of neutrinos on the explosion itself. Similar modeling constraints apply to the current understanding of gravitational wave production in a core collapse. Follow-up observations will be coordinated through SNEWS2.0 \cite{SNEWS:2020tbu}, a global network of neutrino and dark matter detectors likely to detect the first evidence of a nearby supernova. Matched temporal features in the neutrino flux detected across the globe can be used to localize the direction of the supernova with participation from a large number of detectors \cite{Brdar:2018zds, Linzer:2019swe, Coleiro:2020vyj}.

We conclude:
\begin{itemize}
    \item The Rubin LSST survey, DESI, and other cosmic frontier projects should provide target-of-opportunity allocations for the follow-up of rare neutrino (and other multi-messenger) events.
    \item The small-projects portfolio should accommodate support for instrumentation and facilities that provide a full suite of multi-messenger information that supports the science.
    \item Support infrastructure for the real-time optical follow-up of neutrino-triggered events (see \S\ref{tools:sec}).
\end{itemize}

\section{New Projects for Multi-Survey Science With
Rubin Observatory and other Transient Finders}

\textbf{Coordinator(s):} Alex G. Kim
\\
\\
\textbf{Contributor(s):} Greg Aldering, Frederic Courbin, Alex G. Kim, Peter Nugent, Saul Perlmutter, Tommaso Treu 
\\

In the upcoming decade, wide-field imaging surveys, including the Rubin Observatory's LSST, will discover transients that have the potential to probe dark energy and gravity through their 
cosmological distances and motions.
Getting those precision distances and motions
often cannot be obtained from the searches alone but
requires supplemental data from
other targeted follow-up programs.

\subsection{SNe: LS4, PV, Space}

\subsubsection{La Silla Schmidt Southern Survey}

The La Silla Schmidt Southern Survey (LS4) is a 5-year public, wide-field, optical survey using an upgraded 20 square degree QUEST Camera on the ESO Schmidt Telescope at the La Silla Observatory in Chile. It will have first light in late 2022 and use LBNL fully-depleted CCDs to maximize the sensitivity in the optical up to 1\,micron. This survey will complement the {\it Legacy Survey of Space and Time} (LSST) being conducted at the Vera C. Rubin Observatory in two ways. First, it will provide a higher cadence than the LSST over several thousand square degrees of sky each night, allowing a more accurate characterization of brighter and faster evolving transients to 21$^{st}$ magnitude. Second, it will open up a new phase-space for discovery when coupled with the LSST by probing the sky between 12--16$^{th}$ magnitude -- a region where the Rubin Observatory saturates. In addition, a Target of Opportunity (ToO) program will be able to trigger on Multi-Messenger Astronomy (MMA) events with localization uncertainties up to several hundred deg$^2$ in multiple colors very quickly. This project has direct relevance to several cosmology and fundamental physics efforts including: peculiar velocity measurements, and hence fundamental constraints on general relativity, with supernova as standardized candles; gravitational wave standard sirens as probes of the expansion of the Universe and gravity and measurements of the Hubble constant through Type Ia and II-P supernovae.

Why consider a shallow, optical survey in the south at a time during which it will not only overlap with the Rubin observatory, but also with the BlackGEM and DECam facilities? The answer can be broken down into several themes that form the basis of the science case for LS4. The first notion one has to dispel is that the LSST is the do-all and end-all of surveys for transient cosmology and astrophysics. The design of the Rubin Observatory and the LSST is set to achieve several goals in astrophysics, transient science being just one of them.  
\begin{itemize}
    \item {\it The cadence of the LSST WFD survey is not optimal for many transients.} While the reach of the LSST Wide-Fast-Deep survey is impressive, it will leave large gaps in the temporal-color light curves of cosmologically-valuable transients, including spotty early coverage when such transients need to be photometrically-screened as a precursor to spectroscopic follow-up, and gaps in the scientifically important period around peak magnitude due to saturation.
    \item {\it Not all volumes are created equally.} The follow-up capabilities of most of the world's telescopes can only handle the brighter sources discovered by the Rubin Observatory and there is a large swath of transient science in which a timely spectrum is the only path forward for new science. Moreover, nearby peculiar velocity measurements are more accurate. In addition, much of the local universe is inaccessible to the Rubin Observatory due to saturation.
    \item {\it One survey is not the path forward for cosmology and transient astrophysics.} What has become increasingly apparent in astronomy is the power of two or more overlapping surveys. This now forms the backbone of MMA as well as the desire to initiate collaborations between such surveys as Euclid and Roman with the Rubin's LSST, or the DES and DECaLS imaging surveys paving the way for spectroscopy with DESI. 
\end{itemize}

To facilitate the dissemination of new transients discoveries, LS4 will stream its alerts, in near real-time, to all the major transient brokers. Coupled with the Rubin Observatory, the science of both is greatly enhanced. Examples of this include the recent detections of both pre-supernova\cite{2014ApJ...789..104O} and post supernova outbursts\cite{Silverman+13b,GH+17}, in a variety of both core-collapse and thermonuclear supernovae, likely due to mass loss and interaction. While at distances $<$ 100 Mpc, many supernovae saturate for weeks with Rubin, LS4 will accurately observe their lightcurves. Prior to explosion and several years post explosion, Rubin will be sensitive to outbursts at $M_V \leq -11$~mag. These measurements have implications for determining the progenitors of SNe, and hence can help the measurements of the cosmological parameters by removing systematic biases.
Operationally, DOE scientists are well placed to aid in the discovery pipeline work, real-time alert stream as well as triggering and carrying out follow-up for several of the cosmology-focused projects. As the survey compliments several existing DOE Cosmic Frontier projects, such as DESI and LSST-DESC, DOE scientists can play joint roles in each for several of the science goals which overlap with these other surveys.

\subsubsection{Spectroscopic Follow-up of Supernovae for Peculiar Velocities}

Spectroscopic facilities are needed to take advantage of Rubin Observatory supernova discoveries
for peculiar velocity research \cite{snpv}.
Spectra provide transient typing and redshifts
of the
$z \lesssim 0.1$ Type~Ia supernovae discovered within the LSST footprint over the course of its  survey.
Typing (and sub-typing) is required
to determine a supernova's intrinsic luminosity,
which when combined with observed flux yields
its radial distance.
Removing the contribution
of cosmological redshift (inferred from the radial distance) from the observed redshift yields the
supernova peculiar velocity.
This information
cannot be drawn from Rubin data alone and must be drawn from
other facilities.

A peculiar velocity follow-up program obtains for each supernova
an  $R>100$ spectrum to obtain a redshift
from the transient/host light and a transient classification.  The measurement is facilitated
with an integral-field-unit (IFU) spectrograph, from which contributions of the transient and the spatially-structured host galaxy can be distinguished.
The program observes $\sim 10,000$ supernovae per
year up to
$z \sim 0.1$  with magnitudes brighter than $m<21$ over the duration of the LSST Survey.
Access to the survey area of LSST (the southern sky) is essential, though peculiar velocity science would benefit
from additional northern searches and follow-up.

A baseline program that collects the above data can
be scoped based on extrapolations
from the exemplar SNIFS instrument
on the University of Hawaii 88" telescope, which
the Nearby Supernova Factory (SNFactory) experiment used to construct
spectrophotometric light curves of SNe~Ia in the
target redshift range $0.03<z<0.08$.
The limited number of SNe~Ia that explode within the local $z \lesssim0.1$ Universe can be followed with
two 2-m class telescopes instrumented
with IFU spectrgraphs that collectively
monitor the observable extra-Galactic sky.
Incomplete follow-up of half of Rubin discoveries with one telescope
still provides unmatched measurements of
the local peculiar velocity field.

One model for this program
is for DOE scientists 
compete for
instrumentation and survey time on existing telescopes,
or to collaborate with observatory
partners to obtain non-competitive time.
DOE project responsibilities would be in
contributing to the design, development, installation, commissioning, and operations
of the (IFU) spectrograph.
This scope of this activity would fit within
the small projects
portfolio.

\subsubsection{Coordination of Multi-Project SN Cosmology}
As discussed throughout this white paper, a wide variety of transients that can be used as cosmological probes – and that we will discover with the Rubin LSST survey – will lead to significantly more scientific results if we can obtain rapid follow-up spectroscopy and non-optical-band photometry of the transient, ideally within hours or days of the discovery, depending on the timescale of the transient’s progression. Currently, we are already planning for include supernovae, lensed supernovae, and gravitational-wave-triggered optical counterparts, but we also expect other rarer and/or previously uncataloged transients to be discovered when the large surveys turn on. We therefore wish to develop multiple routes to obtaining such spectroscopic and non-optical-band followup, some of which might be directly built and/or paid for, and operated by the DOE and some of which may become available through collaborative agreements.   It will be important for the US HEP collaborations to have flexibility in how they engage with other scientific teams if the best science is to be obtained, and for the DOE to be ready to support some bridging efforts and follow-up efforts that don’t necessarily fall within any one science collaboration so that these efforts to get the best science don’t fall between the cracks.

One particularly important bridging effort which is good to plan for early on relates to the combination of collaborations including several involving space-based telescopes.  The astrophysics and cosmology community has long recognized that many, if not all, of the transients will need a sequence of observations from more than one of Rubin, Roman, Euclid, and ideally JWST, HST, or other space-based assets that can obtain UV-to-near-IR spectroscopy.   Each of these sources of observations has different collaborations that the community will want to bridge.  

We note here that an international space telescope specifically designed around UV-to-near-IR spectrophotometry follow-up of transients is currently being studied, and may well be an additional Rubin/LSST follow-up route.  If this project proceeds, it is likely to be built and operated by philanthropic funding, but there is an opportunity for the US HEP-community to build a new collaboration to lead and manage DOE-mission science with this facility.  Such a collaboration could well position itself to provide observing plans that would optimize the dark-energy science obtained, and build and run data reduction pipelines and archives – and, finally, analyze and publish the cosmology results from these transients. 

Given this positive landscape for the transient science, US-HEP can also take a leadership role by putting in place and supporting a science coordination center that can help bring together the rapidly collecting data from each of the collaborations and enable the on-the-fly follow-up decision making that will need to be made by each separate collaboration concerning which targets to follow next and with what priorities and observing options.   This coordination center would not by itself be part of any of the existing collaborations, although members of them could work together there, but it would make it possible for any or all of them to share real-time observations as appropriate and provide an independent ``neutral'' home for this work.

\subsection{Time Delay Cosmography: US-ELTP and the Time Delay Machine}
\label{tdm}

High cadence high precision monitoring is proposed to carry out precision cosmology with the large number of multiply imaged variable sources that will be discovered in the next decade.

Time delays between multiple images of gravitationally lensed sources such as quasars or supernovae provide an absolute direct distance measurement \citep{TreuMarshall2016,Wong:2020}. With sufficiently large numbers (several hundreds) they can be used to determine the Hubble constant \citep{BirrerTreu:2021} and the cosmic expansion history and the properties of dark energy \citep{Linder:2011}.

The wide field imaging surveys carried out by the Euclid, Roman, and Rubin Observatories will discover thousands of multiply imaged quasars and supernovae \citep{OM10}.  The top100 sample (e.g. 100 quadruply imaged quasars or supernovae with time delays in the range 30-100 days and with deflectors bright enough for detailed kinematics)  will be selected for in depth studies. The larger samples will be available to expand the statistical power of the method based on the lessons learned from the detailed studies.

Recent experience with the Dark Energy Survey \citep{Treu:2018,Lemon:2020} shows that multiply imaged quasars can be discovered in large numbers using ground based imaging data, and confirmed using existing and planned spectroscopic and adaptive optics capabilities \citep{Wizinowich:2020}. The United Sates Extremely Large Telescope Program will be essential to provide deep adaptive-optics assisted spectroscopy to measure spatially resolved stellar kinematics of the deflectors and break the mass sheet degeneracy \citep{BirrerTreu:2021}.
If the Rubin Observatories continues to carry out a time domain survey during its second decade the US-ELTP will also be crucial to take spectra of lensed supernovae and obtain stellar kinematics of the deflectors.

A crucial bottleneck will be the determination of the time delay themselves. The COSMOGRAIL \citep{Courbin:2018} experiment has shown that stability and control over the schedule is a key factor in the success of any monitoring program. While Rubin might deliver 100s of quasar time delays and discovery hundreds of lensed SN over its 10-year life time \cite{OM10,Liao:2015,Goldstein:2019}, only with a dedicated telescope can one  achieve single-season time delays at a few percent precision and build up the top100 sample rapidly enough to achieve breakthroughs in this decade. Control over the schedule is particularly critical to realise the promise of lensed SNe. Lensed SN time delays can be most easily measured in the first few weeks after explosion \cite{Goldstein:2019,Huber:2019}. Typically fainter and with shorter time delays than lensed quasars these targets require an early investment of telescope time to yield few percent precision time-delays. It is therefore important to have the ability to reallocate observing priority to a lensed SN in the rare times that a promising target is live. 

The observational requirements for monitoring are millimag relative precision with daily or quasi-daily cadence, and median image quality of arcsecond or better for deconvolution of blended sources and foreground deflector. In practice, since the bulk of the top100 sources will have i-band magnitudes in the range $i\sim 20-22$, this requires a 3-4m class telescope in a good site, in order to complete the monitoring well within the decade. In terms of instrumentation, the top priority is an optical imager with field of view of $10-30'$ to capture reference stars. A non-thermal infrared channel to the imager would provide additional gains for supernovae light curves.  Second priority in terms of instruments is a optical integral field spectrograph with field of view of 10-30" that would deliver redshifts for the lensed sources (especially the time critical supernovae) and nearby perturbers. Some spectroscopy will be available from surveys like DESI or 4MOST, but a dedicated spectroscopic capability will accelerate the collection of the detailed spectroscopy needed for the study of the top100 sample, and be crucial for real time spectroscopy of lensed supernovae. 

The Time Delay Machine (TDM) experiment can be realized by DOE by re-purposing and managing an existing 3-4m class telescope (or a fraction of one in the North and of one in the South for full hemispheric coverage). In some cases existing instrumentation is sufficient, in others it will have to be built. A non-exhaustive list of telescopes that would be a strong foundation for TDM includes: 4.1m SOAR; 4m Blanco; 4m VISTA; 3.8m UKIRT: 3.5m NTT;
3.5m Galileo; 3.5m Starfire USAF; 3.5m WIYN Arizona; 2.6m VST; 2.6m NOT; 2.2m MPIA; 2$\times$2m LCOGT. A newly built fully robotic telescope would also be excellent of course.

\section{Pan-Experiment Infrastructure}

\textbf{Coordinator(s):} Alex G. Kim 
\\
\\
\textbf{Contributor(s):} Stephen Bailey, Ulysses Braga-Neto, David Jeffery, Alex G. Kim, Gautham Narayan, Lifan Wang     
\\

\subsection{Communication Tools, Data Access, Software}
\label{tools:sec}

Transients as multi-experiment probes necessitate a change in paradigm for how information is shared and analysed between experiments. The Vera Rubin Observatory, Advanced LIGO \& Advanced Virgo, and IceCube will provide a rich array of optical and multi-messenger transient  discoveries. Their full science potential can only be realised with supplemental data from triggered observations at follow-up telescopes. 

The previous generation of surveys were largely self-contained, hosting experiment-specific cyberinfrastructure to share information internally, or externally through annual (or longer) data releases. Transients were announced via Astronomers Telegrams, or the Transient Name Server typically only after the survey team had secured a spectroscopic classification, while transients that were newly discovered in images, but were as yet unclassified, were generally treated as proprietary information and never released. The Zwicky Transient Facility \citep[ZTF,][]{Bellm17} changed this landscape by publicly releasing ``alerts'' -- information on sources that have varied significantly with respect to some reference or ``templat''. These alerts are processed through broker systems such as the Arizona-NOIRLab Transient Alerts and Response to Events System \citep[ANTARES,][]{Narayan18, Matheson21} which allow scientists to execute their own algorithms on these alerts to find new transients from ZTF's public Mid-Scale Innovations Program (MSIP) \emph{in real-time}, scheduling spectroscopic followup within hours with queue-based observatories such as those of the Las Cumbres Observatory. 

In the coming era of multi-messenger astrophysics and the large volume of transient alerts from Rubin, there is a need for generic alert/event layered triggering system that works in real-time laterally and serially \emph{across multiple experiments} to enable swift, efficient decision-making. While individual projects are responsible for identifying and distributing their alerts, alert clients (such as the LSST-DESC) that rely on multi-facility data need processing pipelines that handle the following elements:

\begin{itemize}
\item \textbf{Alerts to targets:}  Broker systems interface with project alert streams and allow users to supply algorithms that identify specific targets of interest. While the broker systems exist, most do not allow users to provide their own algorithms (ANTARES is a notable exception). Efforts such as the Photometric LSST Astronomical Time-series Classification Challenge~\citep[PLAsTiCC,]{Hlozek20, Kessler19} and it's upcoming successor, the Extended LSST Astronomical Time-series Classification Challenge (ELAsTiCC) are preparing the LSST community for the significant challenge posed by automated classification of the entire LSST alert stream, and driving the development of algorithms. However, much work remains to be done to optimize these algorithms \emph{for detecting MMA events using alerts from multiple experiments}, though work has begun to combine alerts from LVK and optical surveys \citep[El-Cid,][]{Chatterjee22} and using only host-galaxy information \citep[GHOST,][]{Gagliano21}.

\item \textbf{Human Assessment:} ``Marshals'' provide a platform by which collaborators can discuss and make decisions about targets. These systems have to be sufficiently general that they can be adapted to meet the needs of data from different surveys, e.g. LVK's GraceDB interface is very different from the YSE-PZ marshal system used for the Young Supernova Experiment~\citep[YSE,][]{Jones21} because the nature of the data is very different. Additionally, survey scientists must be able to modify these components without the additional latency introduced by having to consult with system administrators, suggesting that the infrastructure needs to be containerized as far as possible. Given the importance, fast-evolving nature and rarity of MMA events, these marshals must also be tightly integrated with the communication tools (e.g. Slack) used by survey scientists.  

\item \textbf{Targets to observations:} ``TOM''s combine targets with observing resources to surveys/research groups/individual investigators schedule observations. TOMs can automate the communication of the requested schedule and allow retrieving of the resulting reduced observing products. Increasingly, for the majority of sources, these observations will need to be scheduled automatically as LSST will provide more alerts for common transients than can be feasibly inspected visually, however given the rarity and importance of MMA events, we expect human inspection to always be a crucial step of the analysis of these events.

\item \textbf{Identity and Access Management:} Multi-messenger astrophysics emphasizes the need to combine information from different sources and facilities to glean a complete picture of these enigmatic events, and while the alerts are generally public, value-added information added by surveys (e.g. cross-matching sources against a DESC galaxy catalog with photo-z information) may not be. Similarly, follow-up observations scheduled through TOMs are not generally public. The infrastructure needed to enable multi-experiment probes must a) accommodate users from multiple collaborations, all with their own identity providers, b) make all users aware of the existence of proprietary data sets to avoid redundant observations and c) provide a method to negotiate data rights and share information across different surveys or groups. 

\item \textbf{High-Performance Research Computing:} MMA data is inherently more diverse, heterogeneous and complex than observations from a single survey. Given the low intrinsic rate of these events, relative to normal transients such as supernovae, understanding the population of MMA events with data from multiple experiments will be a complex computational problem. Each experiment has custom tools and pipelines for inference with their data. To understand the demographics of the population fundamentally requires the posterior probability of several competing models, across all MMA events of a specific class like kilonovae, given all of the data. This means that any cyberinfrastructure will have to accommodate each experiment's tools, and provide a unified interface for inference in high-dimensional spaces with this complex data.    
\end{itemize}

The elements of the architecture we describe above are typically considered a ``data lake'' in the parlance of cloud computing, and have typically been called ``research platforms'' within the academe. This kind of integrated cyberinfrastructure implementing a functional pipeline that is useful for Cosmic Frontier science -- a data lake for multi-messenger astrophysics -- requires support for development, training, deployment and operations.
The same cyberinfrastructure developed for the more complex MMA data and used as a research platform by experiments such as LSST DESC will be flexible enough to address the simpler requirements of individual surveys.
Implementing innovative computing cyberinfrastructure will dramatically modernize the paradigm for how scientific research is done in physics and astronomy in the 21st century.

\subsection{Simulation: Digital Twins of Supernovae}

Complete confidence in typical SNe Ia as cosmological distance indicators requires
complete confidence in our theoretical understanding of them. This demands highly realistic explosion models verified by their use in highly realistic radiative transfer calculations that reproduce observed light curve and spectrum data. Though great progress has been made over 6 decades, complete theoretical confidence has not been reached in radiative transfer techniques (i.e., radiative transfer proper and thermal state solution techniques). Recent advancement in statistical treatment of supernova observations especially when combined with deep neural network (NN) is enabling the consolidation of extensive amount of observational data and theoretical models to construct ‘digital twins’ of Type Ia supernovae \cite{Graham15,Stahl:2020MNRAS.496.3553S,Chen:2020ApJS..250...12C,Hu:2022arXiv220202498H}. The digital twins will represent our best knowledge of each individual supernova both observationally and theoretically, which may for the first time allow for the advancement of supernova cosmology from the current purely empirically based status to the stage backed-up by first principle physics. 
The construction of the digital twins requires both high quality observational data and high fidelity theoretical models. Physics-informed neural networks (PINNs) is a new method that may offer significant advantages:  e.g., PINNs treat mixed boundary conditions straightforwardly and they allow redundant constraints such as explicit energy conservation which may speed/enable convergence of the Lambda iteration. PINNs are a new trend in scientific computation, which takes advantage of the considerable computational advancements made by the machine learning community in order to train deep neural networks. There is now a vast high-performance software and hardware infrastructure (e.g., Tensorflow, Google Collab, github) that has been spurred by the deep machine learning revolution, which becomes directly available to PINNs applied to scientific computation problems. PINNs are meshless numerical solvers, which makes them particularly suited to high-dimensional problems in irregular domains, such as radiative transfer modeling \cite{Mishra:2021JQSRT.27007705M}. 

PINNs also have the ability to integrate observational data easily, and can produce predictions with quantified uncertainty by means of Bayesian or ensemble neural network methods. 
In fact, PINNs offer the possibility of solving the theoretical models of each supernova as an inverse problem starting with the observational data and theoretical framework. A digital twin can be constructed by a complete simultaneous solution of the radiative transfer throughout the spacetime domain of SN~Ia emission evolution for each supernova with the composition and kinematics determined to the best the observational data and our theoretical understandings may allow for. 

\subsection{Cosmology Data Repository}
Time-domain events benefit
from external data that contain value-added
information about host galaxies, e.g.~whether a host galaxy spectroscopic
redshift is already known or whether prior photometry from other projects
has seen a transient at this location.  Detailed analysis of the transients may
require access to large amounts of pixel-level image data for ``scene modeling'' while refitting
the per-image point spread function (PSF) and astrometry,
thus requiring more than just on-the-fly postage stamp images around the transient itself.
These external data sources are currently distributed across multiple data archive centers
(e.g.~hosted by NSF, NASA, and ESA), complicating their large-scale and/or real-time joint use.
We advocate support for a cross-site DOE Cosmology Data Repository to facilitate
easy realtime access to the multiple datasets needed to support time-domain analysis
combined with the computing resources necessary to analyze them,
as described in Computational Frontier White Paper ``Data Preservation for Cosmology'' \cite{repository}.

\section{Conclusion}

Projects supported by
the US HEP community
can yield powerful value-added science
through coordination across existing probes and experiments, and modest
investment in new projects that
leverage off of existing ones. In particular, we explored the outstanding potential of standard sirens for dark energy research. We noted that the bottleneck to developing this potential is in the current cosmic surveys and facilities instead of the GW observatories. In particular, to ultimately distinguish between dark energy models, we need to develop the standard sirens methodology fully. In order to do that, we must efficiently detect their electromagnetic counterparts, which requires immediate, deep and wide imaging and spectroscopy. While it is still possible to coordinate this task with current and planned facilities, it will be challenging to meet the requirements to fully develop standard sirens as a significant probe with the proposed telescope facilities.
We advocate support of
\begin{itemize}
    \item Small Projects  ($<\$10$M)
    that get supplemental data
    that enhance the
    science reach of transients discovered by Rubin.
    \item Infrastructure that enables
    cross-experiment, cross-facility coordination and data transfer for
    time-domain astronomical sources.
    \item Theory/modeling that improves
    understanding of the astrophysical
    probes being used to study cosmology.
    \item A US-HEP multi-messenger program, supported with dedicated target-of-opportunity allocations on US-HEP facilities for the follow-up of gravitational wave and rare neutrino events.
    \item Re-purposing of an existing 3-4m class telescope as a dedicated instrument for high-efficiency search and discovery of GW and other transients (e.g. time delays).
    \item The development of a novel standard siren survey program coordinating with the GW observatories to fully incorporate this new observable into our research portfolio for dark energy science. 

\end{itemize}

\section{Acknowledgements}

MESP is funded by the Deutsche Forschungsgemeinschaft (DFG, German Research Foundation) under Germany's Excellence Strategy -- EXC 2121 ``Quantum Universe'' -- 390833306.
AGK is supported by 
the U.S.\ Department of Energy, Office of Science, Office of High Energy 
Physics, under contract No.\ DE-AC02-05CH11231. 
AP acknowledges support for this work was provided by NASA through the NASA Hubble Fellowship grant HST-HF2-51488.001-A awarded by the Space Telescope Science Institute, which is operated by Association of Universities for Research in Astronomy, Inc., for NASA, under contract NAS5-26555.

\bibliographystyle{unsrtnat}
\bibliography{main}

\begin{thebibliography}{87}
\providecommand{\natexlab}[1]{#1}
\providecommand{\url}[1]{\texttt{#1}}
\expandafter\ifx\csname urlstyle\endcsname\relax
  \providecommand{\doi}[1]{doi: #1}\else
  \providecommand{\doi}{doi: \begingroup \urlstyle{rm}\Url}\fi

\bibitem[{Riess} et~al.(1998){Riess}, {Filippenko}, {Challis}, {Clocchiatti},
  {Diercks}, {Garnavich}, {Gilliland}, {Hogan}, {Jha}, {Kirshner},
  {Leibundgut}, {Phillips}, {Reiss}, {Schmidt}, {Schommer}, {Smith},
  {Spyromilio}, {Stubbs}, {Suntzeff}, and {Tonry}]{1998AJ....116.1009R}
Adam~G. {Riess}, Alexei~V. {Filippenko}, Peter {Challis}, Alejandro
  {Clocchiatti}, Alan {Diercks}, Peter~M. {Garnavich}, Ron~L. {Gilliland},
  Craig~J. {Hogan}, Saurabh {Jha}, Robert~P. {Kirshner}, B.~{Leibundgut}, M.~M.
  {Phillips}, David {Reiss}, Brian~P. {Schmidt}, Robert~A. {Schommer}, R.~Chris
  {Smith}, J.~{Spyromilio}, Christopher {Stubbs}, Nicholas~B. {Suntzeff}, and
  John {Tonry}.
\newblock {Observational Evidence from Supernovae for an Accelerating Universe
  and a Cosmological Constant}.
\newblock \emph{\aj}, 116\penalty0 (3):\penalty0 1009--1038, September 1998.
\newblock \doi{10.1086/300499}.

\bibitem[{Perlmutter} et~al.(1999){Perlmutter}, {Aldering}, {Goldhaber},
  et~al.]{1999ApJ...517..565P}
S.~{Perlmutter}, G.~{Aldering}, G.~{Goldhaber}, et~al.
\newblock {Measurements of {\ensuremath{\Omega}} and {\ensuremath{\Lambda}}
  from 42 High-Redshift Supernovae}.
\newblock \emph{\apj}, 517\penalty0 (2):\penalty0 565--586, June 1999.
\newblock \doi{10.1086/307221}.

\bibitem[{Frieman} et~al.(2008){Frieman}, {Turner}, and
  {Huterer}]{2008ARA&A..46..385F}
J.~A. {Frieman}, M.~S. {Turner}, and D.~{Huterer}.
\newblock {Dark energy and the accelerating universe.}
\newblock \emph{\araa}, 46:\penalty0 385--432, September 2008.
\newblock \doi{10.1146/annurev.astro.46.060407.145243}.

\bibitem[{Abdalla} et~al.(2022){Abdalla}, {Abell{\'a}n}, {Aboubrahim},
  et~al.]{2022arXiv220306142A}
Elcio {Abdalla}, Guillermo~Franco {Abell{\'a}n}, Amin {Aboubrahim}, et~al.
\newblock {Cosmology Intertwined: A Review of the Particle Physics,
  Astrophysics, and Cosmology Associated with the Cosmological Tensions and
  Anomalies}.
\newblock \emph{arXiv e-prints}, art. arXiv:2203.06142, March 2022.

\bibitem[{Albrecht} et~al.(2006){Albrecht}, {Bernstein}, {Cahn}, {Freedman},
  {Hewitt}, {Hu}, {Huth}, {Kamionkowski}, {Kolb}, {Knox}, {Mather}, {Staggs},
  and {Suntzeff}]{2006astro.ph..9591A}
Andreas {Albrecht}, Gary {Bernstein}, Robert {Cahn}, Wendy~L. {Freedman},
  Jacqueline {Hewitt}, Wayne {Hu}, John {Huth}, Marc {Kamionkowski}, Edward~W.
  {Kolb}, Lloyd {Knox}, John~C. {Mather}, Suzanne {Staggs}, and Nicholas~B.
  {Suntzeff}.
\newblock {Report of the Dark Energy Task Force}.
\newblock \emph{arXiv e-prints}, art. astro-ph/0609591, September 2006.

\bibitem[{Verde} et~al.(2019){Verde}, {Treu}, and {Riess}]{2019NatAs...3..891V}
Licia {Verde}, Tommaso {Treu}, and Adam~G. {Riess}.
\newblock {Tensions between the early and late Universe}.
\newblock \emph{Nature Astronomy}, 3:\penalty0 891--895, September 2019.
\newblock \doi{10.1038/s41550-019-0902-0}.

\bibitem[{Douspis} et~al.(2019){Douspis}, {Salvati}, and
  {Aghanim}]{2019arXiv190105289D}
Marian {Douspis}, Laura {Salvati}, and Nabila {Aghanim}.
\newblock {On the tension between Large Scale Structures and Cosmic Microwave
  Background}.
\newblock \emph{arXiv e-prints}, art. arXiv:1901.05289, January 2019.

\bibitem[{Baxter} et~al.(2022){Baxter}, {Chang}, {Hearin},
  et~al.]{2022arXiv220306795B}
Eric~J. {Baxter}, Chihway {Chang}, Andrew {Hearin}, et~al.
\newblock {Snowmass2021: Opportunities from Cross-survey Analyses of Static
  Probes}.
\newblock \emph{arXiv e-prints}, art. arXiv:2203.06795, March 2022.

\bibitem[Abbott et~al.(2016)Abbott, Abbott, Abbott, et~al.]{Abbott16}
B.~P. Abbott, R.~Abbott, T.~D. Abbott, et~al.
\newblock Observation of gravitational waves from a binary black hole merger.
\newblock \emph{Phys. Rev. Lett.}, 116:\penalty0 061102, Feb 2016.
\newblock \doi{10.1103/PhysRevLett.116.061102}.
\newblock URL \url{http://link.aps.org/doi/10.1103/PhysRevLett.116.061102}.

\bibitem[{The LIGO Scientific Collaboration} et~al.(2021{\natexlab{a}}){The
  LIGO Scientific Collaboration}, {the Virgo Collaboration}, and {the KAGRA
  Collaboration}]{gwtc3}
{The LIGO Scientific Collaboration}, {the Virgo Collaboration}, and {the KAGRA
  Collaboration}.
\newblock {GWTC-3: Compact Binary Coalescences Observed by LIGO and Virgo
  During the Second Part of the Third Observing Run}.
\newblock \emph{arXiv e-prints}, art. arXiv:2111.03606, November
  2021{\natexlab{a}}.

\bibitem[{Abbott} et~al.(2017{\natexlab{a}}){Abbott}, {Abbott},
  et~al.]{2017PhRvL.119p1101A}
R.~{Abbott}, B.~P.~{Abbott}, T.~D. {Abbott}, et~al.
\newblock {GW170817: Observation of Gravitational Waves from a Binary Neutron
  Star Inspiral}.
\newblock \emph{Phys. Rev. Lett.}, 119\penalty0 (16):\penalty0 161101, October
  2017{\natexlab{a}}.
\newblock \doi{10.1103/PhysRevLett.119.161101}.

\bibitem[{Abbott} et~al.(2017{\natexlab{b}}){Abbott}, {Abbott}, {Abbott},
  {Acernese}, {Ackley}, {Adams}, {Adams}, {Addesso}, {Adhikari}, {Adya}, and
  et~al.]{Abbott17a}
B.~P. {Abbott}, R.~{Abbott}, T.~D. {Abbott}, F.~{Acernese}, K.~{Ackley},
  C.~{Adams}, T.~{Adams}, P.~{Addesso}, R.~X. {Adhikari}, V.~B. {Adya}, and
  et~al.
\newblock {Multi-messenger Observations of a Binary Neutron Star Merger}.
\newblock \emph{Astrophys. J.}, 848:\penalty0 L12, October 2017{\natexlab{b}}.
\newblock \doi{10.3847/2041-8213/aa91c9}.

\bibitem[{Soares-Santos} et~al.(2017){Soares-Santos}, {Holz}, {Annis},
  et~al.]{Soares17}
M.~{Soares-Santos}, D.~E. {Holz}, J.~{Annis}, et~al.
\newblock {The Electromagnetic Counterpart of the Binary Neutron Star Merger
  LIGO/Virgo GW170817. I. Discovery of the Optical Counterpart Using the Dark
  Energy Camera}.
\newblock \emph{Astrophys. J.}, 848:\penalty0 L16, October 2017.
\newblock \doi{10.3847/2041-8213/aa9059}.

\bibitem[{Abbott} et~al.(2017{\natexlab{c}}){Abbott}, {Abbott}, {Abbott},
  {Acernese}, {Ackley}, {Adams}, {Adams}, {Addesso}, {Adhikari}, {Adya}, and
  et~al.]{AbbottH0}
B.~P. {Abbott}, R.~{Abbott}, T.~D. {Abbott}, F.~{Acernese}, K.~{Ackley},
  C.~{Adams}, T.~{Adams}, P.~{Addesso}, R.~X. {Adhikari}, V.~B. {Adya}, and
  et~al.
\newblock {A gravitational-wave standard siren measurement of the Hubble
  constant}.
\newblock \emph{Nature}, 551:\penalty0 85--88, November 2017{\natexlab{c}}.
\newblock \doi{10.1038/nature24471}.

\bibitem[{Soares-Santos} et~al.(2019){Soares-Santos}, {Palmese}, {Hartley},
  et~al.]{2019ApJ...876L...7S}
M.~{Soares-Santos}, A.~{Palmese}, W.~{Hartley}, et~al.
\newblock {First Measurement of the Hubble Constant from a Dark Standard Siren
  using the Dark Energy Survey Galaxies and the LIGO/Virgo Binary-Black-hole
  Merger GW170814}.
\newblock \emph{Astrophys. J.}, 876\penalty0 (1):\penalty0 L7, May 2019.
\newblock \doi{10.3847/2041-8213/ab14f1}.

\bibitem[{Palmese} et~al.(2020){Palmese}, {deVicente}, {Pereira},
  et~al.]{2020ApJ...900L..33P}
A.~{Palmese}, J.~{deVicente}, M.~E.~S. {Pereira}, et~al.
\newblock {A Statistical Standard Siren Measurement of the Hubble Constant from
  the LIGO/Virgo Gravitational Wave Compact Object Merger GW190814 and Dark
  Energy Survey Galaxies}.
\newblock \emph{Astrophys. J.}, 900\penalty0 (2):\penalty0 L33, September 2020.
\newblock \doi{10.3847/2041-8213/abaeff}.

\bibitem[{The LIGO Scientific Collaboration} et~al.(2021{\natexlab{b}}){The
  LIGO Scientific Collaboration}, {the Virgo Collaboration}, and {the KAGRA
  Collaboration}]{2021arXiv211103604T}
{The LIGO Scientific Collaboration}, {the Virgo Collaboration}, and {the KAGRA
  Collaboration}.
\newblock {Constraints on the cosmic expansion history from GWTC-3}.
\newblock \emph{arXiv e-prints}, art. arXiv:2111.03604, November
  2021{\natexlab{b}}.

\bibitem[{Palmese} et~al.(2021){Palmese}, {Bom}, {Mucesh}, and
  {Hartley}]{palmese_StS_DESI}
Antonella {Palmese}, Clecio~R. {Bom}, Sunil {Mucesh}, and William~G. {Hartley}.
\newblock {A standard siren measurement of the Hubble constant using
  gravitational wave events from the first three LIGO/Virgo observing runs and
  the DESI Legacy Survey}.
\newblock \emph{arXiv e-prints}, art. arXiv:2111.06445, November 2021.

\bibitem[{Di Valentino} et~al.(2018){Di Valentino}, {Holz}, {Melchiorri}, and
  {Renzi}]{divalentino2018}
Eleonora {Di Valentino}, Daniel~E. {Holz}, Alessandro {Melchiorri}, and
  Fabrizio {Renzi}.
\newblock {Cosmological impact of future constraints on H$_{0}$ from
  gravitational-wave standard sirens}.
\newblock \emph{\prd}, 98\penalty0 (8):\penalty0 083523, October 2018.
\newblock \doi{10.1103/PhysRevD.98.083523}.

\bibitem[{Ballmer} et~al.(2022){Ballmer}, {Adhikari}, {Badurina}, {Brown},
  {Chattopadhyay}, {Evans}, {Fritschel}, {Hall}, {Hogan}, {Jani}, {Kovachy},
  {Kuns}, {Schwartzman}, {Sigg}, {Slagmolen}, {Vitale}, and
  {Wipf}]{2022arXiv220308228B}
Stefan~W. {Ballmer}, Rana {Adhikari}, Leonardo {Badurina}, Duncan~A. {Brown},
  Swapan {Chattopadhyay}, Matthew {Evans}, Peter {Fritschel}, Evan {Hall},
  Jason~M. {Hogan}, Karan {Jani}, Tim {Kovachy}, Kevin {Kuns}, Ariel
  {Schwartzman}, Daniel {Sigg}, Bram {Slagmolen}, Salvatore {Vitale}, and
  Christopher {Wipf}.
\newblock {Snowmass2021 Cosmic Frontier White Paper: Future Gravitational-Wave
  Detector Facilities}.
\newblock \emph{arXiv e-prints}, art. arXiv:2203.08228, March 2022.

\bibitem[{Herner} et~al.(2020){Herner}, {Annis}, {Brout},
  et~al.]{2020A&C....3300425H}
K.~{Herner}, J.~{Annis}, D.~{Brout}, et~al.
\newblock {Optical follow-up of gravitational wave triggers with DECam during
  the first two LIGO/VIRGO observing runs}.
\newblock \emph{Astronomy and Computing}, 33:\penalty0 100425, October 2020.
\newblock \doi{10.1016/j.ascom.2020.100425}.

\bibitem[{Garcia} et~al.(2020){Garcia}, {Morgan}, {Herner},
  et~al.]{2020ApJ...903...75G}
A.~{Garcia}, R.~{Morgan}, K.~{Herner}, et~al.
\newblock {A DESGW Search for the Electromagnetic Counterpart to the LIGO/Virgo
  Gravitational-wave Binary Neutron Star Merger Candidate S190510g}.
\newblock \emph{\apj}, 903\penalty0 (1):\penalty0 75, November 2020.
\newblock \doi{10.3847/1538-4357/abb823}.

\bibitem[{Morgan} et~al.(2020){Morgan}, {Soares-Santos}, {Annis},
  et~al.]{2020ApJ...901...83M}
R.~{Morgan}, M.~{Soares-Santos}, J.~{Annis}, et~al.
\newblock {Constraints on the Physical Properties of GW190814 through
  Simulations Based on DECam Follow-up Observations by the Dark Energy Survey}.
\newblock \emph{\apj}, 901\penalty0 (1):\penalty0 83, September 2020.
\newblock \doi{10.3847/1538-4357/abafaa}.

\bibitem[{Shandonay} et~al.(2022){Shandonay}, {Morgan}, {Bechtol}, {Bom},
  {Nord}, {Garcia}, {Henghes}, {Herner}, {Tabbutt}, {Palmese}, {Santana-Silva},
  {Soares-Santos}, {Gill}, and {Garc{\'\i}a-Bellido}]{2022ApJ...925...44S}
A.~{Shandonay}, R.~{Morgan}, K.~{Bechtol}, C.~R. {Bom}, B.~{Nord}, A.~{Garcia},
  B.~{Henghes}, K.~{Herner}, M.~{Tabbutt}, A.~{Palmese}, L.~{Santana-Silva},
  M.~{Soares-Santos}, M.~S.~S. {Gill}, and J.~{Garc{\'\i}a-Bellido}.
\newblock {Expediting DECam Multimessenger Counterpart Searches with
  Convolutional Neural Networks}.
\newblock \emph{\apj}, 925\penalty0 (1):\penalty0 44, January 2022.
\newblock \doi{10.3847/1538-4357/ac3760}.

\bibitem[{Tucker} et~al.(2022){Tucker}, {Wiesner}, {Allam},
  et~al.]{2022ApJ...929..115T}
D.~L. {Tucker}, M.~P. {Wiesner}, S.~S. {Allam}, et~al.
\newblock {SOAR/Goodman Spectroscopic Assessment of Candidate Counterparts of
  the LIGO/Virgo Event GW190814}.
\newblock \emph{\apj}, 929\penalty0 (2):\penalty0 115, April 2022.
\newblock \doi{10.3847/1538-4357/ac5b60}.

\bibitem[{Bom} et~al.(2022){Bom}, {Annis}, {Garcia}, and et~al.]{Bom2022}
C.~R. {Bom}, J.~{Annis}, A.~{Garcia}, and et~al.
\newblock {Designing an Optimal Kilonovae Search using DECam for Gravitational
  Wave Events}.
\newblock in preparation, 2022.

\bibitem[Collaboration et~al.(2012)]{lsst2012large}
LSST Dark Energy~Science Collaboration et~al.
\newblock Large synoptic survey telescope: dark energy science collaboration.
\newblock \emph{arXiv preprint arXiv:1211.0310}, 2012.

\bibitem[{Borhanian} and {Sathyaprakash}(2022)]{borhanian2022listening}
Ssohrab {Borhanian} and B.~S. {Sathyaprakash}.
\newblock {Listening to the Universe with Next Generation Ground-Based
  Gravitational-Wave Detectors}.
\newblock \emph{arXiv e-prints}, art. arXiv:2202.11048, February 2022.

\bibitem[{Planck Collaboration} et~al.(2020){Planck Collaboration}, {Aghanim},
  {Akrami}, et~al.]{2020A&A...641A...6P}
{Planck Collaboration}, N.~{Aghanim}, Y.~{Akrami}, et~al.
\newblock {Planck 2018 results. VI. Cosmological parameters}.
\newblock \emph{\aap}, 641:\penalty0 A6, September 2020.
\newblock \doi{10.1051/0004-6361/201833910}.

\bibitem[{Planck Collaboration} et~al.(2016){Planck Collaboration}, {Ade},
  {Aghanim}, et~al.]{2016A&A...594A..14P}
{Planck Collaboration}, P.~A.~R. {Ade}, N.~{Aghanim}, et~al.
\newblock {Planck 2015 results. XIV. Dark energy and modified gravity}.
\newblock \emph{\aap}, 594:\penalty0 A14, September 2016.
\newblock \doi{10.1051/0004-6361/201525814}.

\bibitem[{Shafieloo} et~al.(2020){Shafieloo}, {Keeley}, and
  {Linder}]{Shafieloo2018}
Arman {Shafieloo}, Ryan~E. {Keeley}, and Eric~V. {Linder}.
\newblock {Will cosmic gravitational wave sirens determine the Hubble
  constant?}
\newblock \emph{\jcap}, 2020\penalty0 (3):\penalty0 019, March 2020.
\newblock \doi{10.1088/1475-7516/2020/03/019}.

\bibitem[Palmese et~al.(2021)Palmese, Fishbach, Burke, Annis, and
  Liu]{Palmese_2021}
A.~Palmese, M.~Fishbach, C.~J. Burke, J.~Annis, and X.~Liu.
\newblock Do ligo/virgo black hole mergers produce agn flares? the case of
  gw190521 and prospects for reaching a confident association.
\newblock \emph{The Astrophysical Journal Letters}, 914\penalty0 (2):\penalty0
  L34, Jun 2021.
\newblock ISSN 2041-8213.
\newblock \doi{10.3847/2041-8213/ac0883}.
\newblock URL \url{http://dx.doi.org/10.3847/2041-8213/ac0883}.

\bibitem[{Palmese} and {Kim}(2020)]{palmese20}
Antonella {Palmese} and Alex~G. {Kim}.
\newblock {Probing gravity and growth of structure with gravitational waves and
  galaxies' peculiar velocity}.
\newblock \emph{arXiv e-prints}, May 2020.

\bibitem[Speri et~al.(2021)Speri, Tamanini, Caldwell, Gair, and
  Wang]{speri2021}
Lorenzo Speri, Nicola Tamanini, Robert~R. Caldwell, Jonathan~R. Gair, and
  Benjamin Wang.
\newblock Testing the quasar hubble diagram with lisa standard sirens.
\newblock \emph{Physical Review D}, 103\penalty0 (8), Apr 2021.
\newblock ISSN 2470-0029.
\newblock \doi{10.1103/physrevd.103.083526}.
\newblock URL \url{http://dx.doi.org/10.1103/PhysRevD.103.083526}.

\bibitem[{Palmese} et~al.(2019){Palmese}, {Graur}, {Annis}, et~al.]{palmese_WP}
Antonella {Palmese}, Or~{Graur}, James~T. {Annis}, et~al.
\newblock {Gravitational wave cosmology and astrophysics with large
  spectroscopic galaxy surveys}.
\newblock \emph{\baas}, 51\penalty0 (3):\penalty0 310, May 2019.

\bibitem[Guidorzi et~al.(2017)Guidorzi, Margutti, Brout, Scolnic, Fong,
  Alexander, Cowperthwaite, Annis, Berger, Blanchard, Chornock, Coppejans,
  Eftekhari, Frieman, Huterer, Nicholl, Soares-Santos, Terreran, Villar, and
  Williams]{guidorzi}
C.~Guidorzi, R.~Margutti, D.~Brout, D.~Scolnic, W.~Fong, K.~D. Alexander, P.~S.
  Cowperthwaite, J.~Annis, E.~Berger, P.~K. Blanchard, R.~Chornock, D.~L.
  Coppejans, T.~Eftekhari, J.~A. Frieman, D.~Huterer, M.~Nicholl,
  M.~Soares-Santos, G.~Terreran, V.~A. Villar, and P.~K.~G. Williams.
\newblock Improved constraints on h 0 from a combined analysis of
  gravitational-wave and electromagnetic emission from gw170817.
\newblock \emph{The Astrophysical Journal}, 851\penalty0 (2):\penalty0 L36, Dec
  2017.
\newblock ISSN 2041-8213.
\newblock \doi{10.3847/2041-8213/aaa009}.
\newblock URL \url{http://dx.doi.org/10.3847/2041-8213/aaa009}.

\bibitem[Dhawan et~al.(2020)Dhawan, Bulla, Goobar, Sagués~Carracedo, and
  Setzer]{KNH02020}
S.~Dhawan, M.~Bulla, A.~Goobar, A.~Sagués~Carracedo, and C.~N. Setzer.
\newblock Constraining the observer angle of the kilonova at2017gfo associated
  with gw170817: Implications for the hubble constant.
\newblock \emph{The Astrophysical Journal}, 888\penalty0 (2):\penalty0 67, Jan
  2020.
\newblock ISSN 1538-4357.
\newblock \doi{10.3847/1538-4357/ab5799}.
\newblock URL \url{http://dx.doi.org/10.3847/1538-4357/ab5799}.

\bibitem[Ahlers et~al.(2022)Ahlers, Albert, Allen, et~al.]{Engel:2022}
M.~Ahlers, A.~Albert, A.~Allen, et~al.
\newblock Advancing the landscape of multimessenger science in the next decade.
\newblock \emph{Snowmass2021}, 2022.

\bibitem[Ando et~al.(2022)]{Ando:2022}
S.~Ando et~al.
\newblock Snowmass2021 cosmic frontier: Synergies between dark matter searches
  and multiwavelength/multimessenger astrophysics.
\newblock \emph{Snowmass2021}, 2022.

\bibitem[Morgan et~al.(2019)]{DES:2019hqa}
R.~Morgan et~al.
\newblock {A DECam Search for Explosive Optical Transients Associated with
  IceCube Neutrinos}.
\newblock \emph{Astrophys. J.}, 883:\penalty0 125, 2019.
\newblock \doi{10.3847/1538-4357/ab3a45}.

\bibitem[Stein et~al.(2021)]{Stein:2020xhk}
Robert Stein et~al.
\newblock {A tidal disruption event coincident with a high-energy neutrino}.
\newblock \emph{Nature Astron.}, 5\penalty0 (5):\penalty0 510--518, 2021.
\newblock \doi{10.1038/s41550-020-01295-8}.

\bibitem[{Palmese} et~al.(2021){Palmese}, {BenZvi}, {Bailey}, {Davis}, {Kim},
  {Landriau}, {Moutard}, {Myers}, {Nugent}, {Raichoor}, {Schlafly}, {Schlegel},
  {Demirbozan}, {Della Costa}, {Transients}, and {Low-z Cosmology Working
  Group}]{2021GCN.30923....1P}
Antonella {Palmese}, Segev {BenZvi}, Stephen {Bailey}, Tamara {Davis}, Alex
  {Kim}, Martin {Landriau}, David {Moutard}, Adam {Myers}, Peter {Nugent},
  Anand {Raichoor}, Edward {Schlafly}, David {Schlegel}, Umut {Demirbozan},
  John {Della Costa}, the~DESI {Transients}, and {Low-z Cosmology Working
  Group}.
\newblock {IceCube-210922A: DESI Observations}.
\newblock \emph{GRB Coordinates Network}, 30923:\penalty0 1, October 2021.

\bibitem[Collaboration(2021)]{IceCube:210922A}
IceCube Collaboration.
\newblock {IceCube-210922A - IceCube observation of a high-energy neutrino
  candidate track-like event}.
\newblock \url{https://gcn.gsfc.nasa.gov/other/icecube_210922A.gcn3}, 2021.

\bibitem[Duan et~al.(2010)Duan, Fuller, and Qian]{Duan:2010bg}
Huaiyu Duan, George~M. Fuller, and Yong-Zhong Qian.
\newblock {Collective Neutrino Oscillations}.
\newblock \emph{Ann. Rev. Nucl. Part. Sci.}, 60:\penalty0 569--594, 2010.
\newblock \doi{10.1146/annurev.nucl.012809.104524}.

\bibitem[Sasaki et~al.(2020)Sasaki, Takiwaki, Kawagoe, Horiuchi, and
  Ishidoshiro]{Sasaki:2019jny}
Hirokazu Sasaki, Tomoya Takiwaki, Shio Kawagoe, Shunsaku Horiuchi, and Koji
  Ishidoshiro.
\newblock {Detectability of Collective Neutrino Oscillation Signatures in the
  Supernova Explosion of a 8.8 $M_\odot$ star}.
\newblock \emph{Phys. Rev. D}, 101\penalty0 (6):\penalty0 063027, 2020.
\newblock \doi{10.1103/PhysRevD.101.063027}.

\bibitem[Berryman et~al.(2022)Berryman, Blinov, Brdar, et~al.]{Berryman:2022}
J.~M. Berryman, N.~Blinov, V.~Brdar, et~al.
\newblock Neutrino self-interactions: A white paper.
\newblock \emph{Snowmass2021}, 2022.

\bibitem[Fischer et~al.(2016)Fischer, Chakraborty, Giannotti, Mirizzi, Payez,
  and Ringwald]{Fischer:2016cyd}
Tobias Fischer, Sovan Chakraborty, Maurizio Giannotti, Alessandro Mirizzi,
  Alexandre Payez, and Andreas Ringwald.
\newblock {Probing axions with the neutrino signal from the next galactic
  supernova}.
\newblock \emph{Phys. Rev. D}, 94\penalty0 (8):\penalty0 085012, 2016.
\newblock \doi{10.1103/PhysRevD.94.085012}.

\bibitem[M\"uller(2019)]{Muller:2019upo}
B.~M\"uller.
\newblock {Neutrino Emission as Diagnostics of Core-Collapse Supernovae}.
\newblock \emph{Ann. Rev. Nucl. Part. Sci.}, 69:\penalty0 253--278, 2019.
\newblock \doi{10.1146/annurev-nucl-101918-023434}.

\bibitem[Franarin et~al.(2018)Franarin, Davis, and Fairbairn]{Franarin:2017jnd}
Tarso Franarin, Jonathan~H. Davis, and Malcolm Fairbairn.
\newblock {Prospects for detecting eV-scale sterile neutrinos from a galactic
  supernova}.
\newblock \emph{JCAP}, 09:\penalty0 002, 2018.
\newblock \doi{10.1088/1475-7516/2018/09/002}.

\bibitem[Arg\"uelles et~al.(2019)Arg\"uelles, Brdar, and
  Kopp]{Arguelles:2016uwb}
Carlos~A. Arg\"uelles, Vedran Brdar, and Joachim Kopp.
\newblock {Production of keV Sterile Neutrinos in Supernovae: New Constraints
  and Gamma Ray Observables}.
\newblock \emph{Phys. Rev. D}, 99\penalty0 (4):\penalty0 043012, 2019.
\newblock \doi{10.1103/PhysRevD.99.043012}.

\bibitem[Nishizawa and Nakamura(2014)]{Nishizawa:2014zna}
Atsushi Nishizawa and Takashi Nakamura.
\newblock {Measuring Speed of Gravitational Waves by Observations of Photons
  and Neutrinos from Compact Binary Mergers and Supernovae}.
\newblock \emph{Phys. Rev. D}, 90\penalty0 (4):\penalty0 044048, 2014.
\newblock \doi{10.1103/PhysRevD.90.044048}.

\bibitem[Pagliaroli and Vissani(2011)]{Pagliaroli:2011zz}
G.~Pagliaroli and F.~Vissani.
\newblock {Supernova neutrinos and gravitational waves}.
\newblock \emph{Nucl. Phys. B Proc. Suppl.}, 217:\penalty0 278--280, 2011.
\newblock \doi{10.1016/j.nuclphysbps.2011.04.119}.

\bibitem[Nakamura et~al.(2016)Nakamura, Horiuchi, Tanaka, Hayama, Takiwaki, and
  Kotake]{Nakamura:2016kkl}
Ko~Nakamura, Shunsaku Horiuchi, Masaomi Tanaka, Kazuhiro Hayama, Tomoya
  Takiwaki, and Kei Kotake.
\newblock {Multimessenger signals of long-term core-collapse supernova
  simulations: synergetic observation strategies}.
\newblock \emph{Mon. Not. Roy. Astron. Soc.}, 461\penalty0 (3):\penalty0
  3296--3313, 2016.
\newblock \doi{10.1093/mnras/stw1453}.

\bibitem[Al~Kharusi et~al.(2021)]{SNEWS:2020tbu}
S.~Al~Kharusi et~al.
\newblock {SNEWS 2.0: a next-generation supernova early warning system for
  multi-messenger astronomy}.
\newblock \emph{New J. Phys.}, 23\penalty0 (3):\penalty0 031201, 2021.
\newblock \doi{10.1088/1367-2630/abde33}.

\bibitem[Brdar et~al.(2018)Brdar, Lindner, and Xu]{Brdar:2018zds}
Vedran Brdar, Manfred Lindner, and Xun-Jie Xu.
\newblock {Neutrino astronomy with supernova neutrinos}.
\newblock \emph{JCAP}, 04:\penalty0 025, 2018.
\newblock \doi{10.1088/1475-7516/2018/04/025}.

\bibitem[Linzer and Scholberg(2019)]{Linzer:2019swe}
N.~B. Linzer and K.~Scholberg.
\newblock {Triangulation Pointing to Core-Collapse Supernovae with
  Next-Generation Neutrino Detectors}.
\newblock \emph{Phys. Rev. D}, 100\penalty0 (10):\penalty0 103005, 2019.
\newblock \doi{10.1103/PhysRevD.100.103005}.

\bibitem[Coleiro et~al.(2020)Coleiro, Colomer~Molla, Dornic, Lincetto, and
  Kulikovskiy]{Coleiro:2020vyj}
A.~Coleiro, M.~Colomer~Molla, D.~Dornic, M.~Lincetto, and V.~Kulikovskiy.
\newblock {Combining neutrino experimental light-curves for pointing to the
  next galactic core-collapse supernova}.
\newblock \emph{Eur. Phys. J. C}, 80\penalty0 (9):\penalty0 856, 2020.
\newblock \doi{10.1140/epjc/s10052-020-8407-7}.

\bibitem[{Ofek} et~al.(2014){Ofek}, {Sullivan}, {Shaviv},
  et~al.]{2014ApJ...789..104O}
Eran~O. {Ofek}, Mark {Sullivan}, Nir~J. {Shaviv}, et~al.
\newblock {Precursors Prior to Type IIn Supernova Explosions are Common:
  Precursor Rates, Properties, and Correlations}.
\newblock \emph{\apj}, 789\penalty0 (2):\penalty0 104, July 2014.
\newblock \doi{10.1088/0004-637X/789/2/104}.

\bibitem[{Silverman} et~al.(2013){Silverman}, {Nugent}, {Gal-Yam}, {Sullivan},
  {Howell}, {Filippenko}, {Pan}, {Cenko}, and {Hook}]{Silverman+13b}
J.~M. {Silverman}, P.~E. {Nugent}, A.~{Gal-Yam}, M.~{Sullivan}, D.~A. {Howell},
  A.~V. {Filippenko}, Y.-C. {Pan}, S.~B. {Cenko}, and I.~M. {Hook}.
\newblock {Late-time Spectral Observations of the Strongly Interacting Type Ia
  Supernova PTF11kx}.
\newblock \emph{\apj}, 772:\penalty0 125, August 2013.
\newblock \doi{10.1088/0004-637X/772/2/125}.

\bibitem[{Graham} et~al.(2017){Graham}, {Harris}, {Fox}, {Nugent}, {Kasen},
  {Silverman}, and {Filippenko}]{GH+17}
M.~L. {Graham}, C.~E. {Harris}, O.~D. {Fox}, P.~E. {Nugent}, D.~{Kasen}, J.~M.
  {Silverman}, and A.~V. {Filippenko}.
\newblock {PTF11kx: A Type Ia Supernova with Hydrogen Emission Persisting after
  3.5 Years}.
\newblock \emph{\apj}, 843:\penalty0 102, July 2017.
\newblock \doi{10.3847/1538-4357/aa78ee}.

\bibitem[Kim(2021)]{snpv}
A.~Kim.
\newblock Probing gravity with type ia supernova peculiar velocities, 2021.
\newblock Snowmass2021: Letter of Interest.

\bibitem[{Treu} and {Marshall}(2016)]{TreuMarshall2016}
Tommaso {Treu} and Philip~J. {Marshall}.
\newblock {Time delay cosmography}.
\newblock \emph{\aapr}, 24\penalty0 (1):\penalty0 11, July 2016.
\newblock \doi{10.1007/s00159-016-0096-8}.

\bibitem[{Wong} et~al.(2020){Wong}, {Suyu}, {Chen}, {Rusu}, {Millon}, {Sluse},
  {Bonvin}, {Fassnacht}, {Taubenberger}, {Auger}, {Birrer}, {Chan}, {Courbin},
  {Hilbert}, {Tihhonova}, {Treu}, {Agnello}, {Ding}, {Jee}, {Komatsu},
  {Shajib}, {Sonnenfeld}, {Bland ford}, {Koopmans}, {Marshall}, and
  {Meylan}]{Wong:2020}
Kenneth~C. {Wong}, Sherry~H. {Suyu}, Geoff C.~F. {Chen}, Cristian~E. {Rusu},
  Martin {Millon}, Dominique {Sluse}, Vivien {Bonvin}, Christopher~D.
  {Fassnacht}, Stefan {Taubenberger}, Matthew~W. {Auger}, Simon {Birrer}, James
  H.~H. {Chan}, Frederic {Courbin}, Stefan {Hilbert}, Olga {Tihhonova}, Tommaso
  {Treu}, Adriano {Agnello}, Xuheng {Ding}, Inh {Jee}, Eiichiro {Komatsu},
  Anowar~J. {Shajib}, Alessandro {Sonnenfeld}, Roger~D. {Bland ford}, L{\'e}on
  V.~E. {Koopmans}, Philip~J. {Marshall}, and Georges {Meylan}.
\newblock {H0LiCOW XIII. A 2.4\% measurement of H$_{0}$ from lensed quasars:
  5.3{\ensuremath{\sigma}} tension between early and late-Universe probes}.
\newblock \emph{\mnras}, June 2020.
\newblock \doi{10.1093/mnras/stz3094}.

\bibitem[{Birrer} and {Treu}(2021)]{BirrerTreu:2021}
Simon {Birrer} and Tommaso {Treu}.
\newblock {TDCOSMO. V. Strategies for precise and accurate measurements of the
  Hubble constant with strong lensing}.
\newblock \emph{\aap}, 649:\penalty0 A61, May 2021.
\newblock \doi{10.1051/0004-6361/202039179}.

\bibitem[{Linder}(2011)]{Linder:2011}
Eric~V. {Linder}.
\newblock {Lensing time delays and cosmological complementarity}.
\newblock \emph{\prd}, 84\penalty0 (12):\penalty0 123529, December 2011.
\newblock \doi{10.1103/PhysRevD.84.123529}.

\bibitem[{Oguri} and {Marshall}(2010)]{OM10}
Masamune {Oguri} and Philip~J. {Marshall}.
\newblock {Gravitationally lensed quasars and supernovae in future wide-field
  optical imaging surveys}.
\newblock \emph{\mnras}, 405\penalty0 (4):\penalty0 2579--2593, July 2010.
\newblock \doi{10.1111/j.1365-2966.2010.16639.x}.

\bibitem[{Treu} et~al.(2018){Treu}, {Agnello}, {Baumer}, et~al.]{Treu:2018}
T.~{Treu}, A.~{Agnello}, M.~A. {Baumer}, et~al.
\newblock {The STRong lensing Insights into the Dark Energy Survey (STRIDES)
  2016 follow-up campaign - I. Overview and classification of candidates
  selected by two techniques}.
\newblock \emph{\mnras}, 481\penalty0 (1):\penalty0 1041--1054, November 2018.
\newblock \doi{10.1093/mnras/sty2329}.

\bibitem[{Lemon} et~al.(2020){Lemon}, {Auger}, {McMahon}, et~al.]{Lemon:2020}
C.~{Lemon}, M.~W. {Auger}, R.~{McMahon}, et~al.
\newblock {The STRong lensing Insights into the Dark Energy Survey (STRIDES)
  2017/2018 follow-up campaign: discovery of 10 lensed quasars and 10 quasar
  pairs}.
\newblock \emph{\mnras}, 494\penalty0 (3):\penalty0 3491--3511, March 2020.
\newblock \doi{10.1093/mnras/staa652}.

\bibitem[{Wizinowich} et~al.(2020){Wizinowich}, {Chin}, {Correia}, {Lu},
  {Brown}, {Casey}, {Cetre}, {Delorme}, {Gers}, {Hunter}, {Lilley}, {Ragland},
  {Surendran}, {Wetherell}, {Ghez}, {Do}, {Jones}, {Liu}, {Mawet}, {Max},
  {Morris}, {Treu}, and {Wright}]{Wizinowich:2020}
P.~{Wizinowich}, J.~{Chin}, C.~{Correia}, J.~{Lu}, T.~{Brown}, K.~{Casey},
  S.~{Cetre}, J.~R. {Delorme}, L.~{Gers}, L.~{Hunter}, S.~{Lilley},
  S.~{Ragland}, A.~{Surendran}, E.~{Wetherell}, A.~{Ghez}, T.~{Do}, T.~{Jones},
  M.~{Liu}, D.~{Mawet}, C.~{Max}, M.~{Morris}, T.~{Treu}, and S.~{Wright}.
\newblock {Keck all sky precision adaptive optics}.
\newblock In \emph{Society of Photo-Optical Instrumentation Engineers (SPIE)
  Conference Series}, volume 11448 of \emph{Society of Photo-Optical
  Instrumentation Engineers (SPIE) Conference Series}, page 114480E, December
  2020.
\newblock \doi{10.1117/12.2560017}.

\bibitem[{Courbin} et~al.(2018){Courbin}, {Bonvin}, {Buckley-Geer},
  et~al.]{Courbin:2018}
F.~{Courbin}, V.~{Bonvin}, E.~{Buckley-Geer}, et~al.
\newblock {COSMOGRAIL: the COSmological MOnitoring of GRAvItational Lenses.
  XVI. Time delays for the quadruply imaged quasar DES J0408-5354 with
  high-cadence photometric monitoring}.
\newblock \emph{\aap}, 609:\penalty0 A71, January 2018.
\newblock \doi{10.1051/0004-6361/201731461}.

\bibitem[{Liao} et~al.(2015){Liao}, {Treu}, {Marshall}, {Fassnacht},
  {Rumbaugh}, {Dobler}, {Aghamousa}, {Bonvin}, {Courbin}, {Hojjati}, {Jackson},
  {Kashyap}, {Rathna Kumar}, {Linder}, {Mandel}, {Meng}, {Meylan}, {Moustakas},
  {Prabhu}, {Romero-Wolf}, {Shafieloo}, {Siemiginowska}, {Stalin}, {Tak},
  {Tewes}, and {van Dyk}]{Liao:2015}
Kai {Liao}, Tommaso {Treu}, Phil {Marshall}, Christopher~D. {Fassnacht}, Nick
  {Rumbaugh}, Gregory {Dobler}, Amir {Aghamousa}, Vivien {Bonvin}, Frederic
  {Courbin}, Alireza {Hojjati}, Neal {Jackson}, Vinay {Kashyap}, S.~{Rathna
  Kumar}, Eric {Linder}, Kaisey {Mandel}, Xiao-Li {Meng}, Georges {Meylan},
  Leonidas~A. {Moustakas}, Tushar~P. {Prabhu}, Andrew {Romero-Wolf}, Arman
  {Shafieloo}, Aneta {Siemiginowska}, Chelliah~S. {Stalin}, Hyungsuk {Tak},
  Malte {Tewes}, and David {van Dyk}.
\newblock {Strong Lens Time Delay Challenge. II. Results of TDC1}.
\newblock \emph{\apj}, 800\penalty0 (1):\penalty0 11, February 2015.
\newblock \doi{10.1088/0004-637X/800/1/11}.

\bibitem[{Goldstein} et~al.(2019){Goldstein}, {Nugent}, and
  {Goobar}]{Goldstein:2019}
Daniel~A. {Goldstein}, Peter~E. {Nugent}, and Ariel {Goobar}.
\newblock {Rates and Properties of Supernovae Strongly Gravitationally Lensed
  by Elliptical Galaxies in Time-domain Imaging Surveys}.
\newblock \emph{\apjs}, 243\penalty0 (1):\penalty0 6, July 2019.
\newblock \doi{10.3847/1538-4365/ab1fe0}.

\bibitem[{Huber} et~al.(2019){Huber}, {Suyu}, {Noebauer}, {Bonvin},
  {Rothchild}, {Chan}, {Awan}, {Courbin}, {Kromer}, {Marshall}, {Oguri},
  {Ribeiro}, and {LSST Dark Energy Science Collaboration}]{Huber:2019}
S.~{Huber}, S.~H. {Suyu}, U.~M. {Noebauer}, V.~{Bonvin}, D.~{Rothchild},
  J.~H.~H. {Chan}, H.~{Awan}, F.~{Courbin}, M.~{Kromer}, P.~{Marshall},
  M.~{Oguri}, T.~{Ribeiro}, and {LSST Dark Energy Science Collaboration}.
\newblock {Strongly lensed SNe Ia in the era of LSST: observing cadence for
  lens discoveries and time-delay measurements}.
\newblock \emph{\aap}, 631:\penalty0 A161, November 2019.
\newblock \doi{10.1051/0004-6361/201935370}.

\bibitem[{Bellm} and {Kulkarni}(2017)]{Bellm17}
E.~{Bellm} and S.~{Kulkarni}.
\newblock {The unblinking eye on the sky}.
\newblock \emph{Nature Astronomy}, 1:\penalty0 0071, March 2017.
\newblock \doi{10.1038/s41550-017-0071}.

\bibitem[{Narayan} et~al.(2018){Narayan}, {Zaidi}, {Soraisam}, {Wang},
  {Lochner}, {Matheson}, {Saha}, {Yang}, {Zhao}, {Kececioglu}, {Scheidegger},
  {Snodgrass}, {Axelrod}, {Jenness}, {Maier}, {Ridgway}, {Seaman}, {Evans},
  {Singh}, {Taylor}, {Toeniskoetter}, {Welch}, {Zhu}, and {ANTARES
  Collaboration}]{Narayan18}
Gautham {Narayan}, Tayeb {Zaidi}, Monika~D. {Soraisam}, Zhe {Wang}, Michelle
  {Lochner}, Thomas {Matheson}, Abhijit {Saha}, Shuo {Yang}, Zhenge {Zhao},
  John {Kececioglu}, Carlos {Scheidegger}, Richard~T. {Snodgrass}, Tim
  {Axelrod}, Tim {Jenness}, Robert~S. {Maier}, Stephen~T. {Ridgway}, Robert~L.
  {Seaman}, Eric~Michael {Evans}, Navdeep {Singh}, Clark {Taylor}, Jackson
  {Toeniskoetter}, Eric {Welch}, Songzhe {Zhu}, and {ANTARES Collaboration}.
\newblock {Machine-learning-based Brokers for Real-time Classification of the
  LSST Alert Stream}.
\newblock \emph{\apjs}, 236\penalty0 (1):\penalty0 9, May 2018.
\newblock \doi{10.3847/1538-4365/aab781}.

\bibitem[{Matheson} et~al.(2021){Matheson}, {Stubens}, {Wolf}, {Lee},
  {Narayan}, {Saha}, {Scott}, {Soraisam}, {Bolton}, {Hauger}, {Silva},
  {Kececioglu}, {Scheidegger}, {Snodgrass}, {Aleo}, {Evans-Jacquez}, {Singh},
  {Wang}, {Yang}, and {Zhao}]{Matheson21}
Thomas {Matheson}, Carl {Stubens}, Nicholas {Wolf}, Chien-Hsiu {Lee}, Gautham
  {Narayan}, Abhijit {Saha}, Adam {Scott}, Monika {Soraisam}, Adam~S. {Bolton},
  Benjamin {Hauger}, David~R. {Silva}, John {Kececioglu}, Carlos {Scheidegger},
  Richard {Snodgrass}, Patrick~D. {Aleo}, Eric {Evans-Jacquez}, Navdeep
  {Singh}, Zhe {Wang}, Shuo {Yang}, and Zhenge {Zhao}.
\newblock {The ANTARES Astronomical Time-domain Event Broker}.
\newblock \emph{\aj}, 161\penalty0 (3):\penalty0 107, March 2021.
\newblock \doi{10.3847/1538-3881/abd703}.

\bibitem[{Hlo{\v{z}}ek} et~al.(2020){Hlo{\v{z}}ek}, {Ponder}, {Malz}, {Dai},
  {Narayan}, {Ishida}, {Allam}, {Bahmanyar}, {Biswas}, {Galbany}, {Jha},
  {Jones}, {Kessler}, {Lochner}, {Mahabal}, {Mandel}, {Mart{\'\i}nez-Galarza},
  {McEwen}, {Muthukrishna}, {Peiris}, {Peters}, and {Setzer}]{Hlozek20}
R.~{Hlo{\v{z}}ek}, K.~A. {Ponder}, A.~I. {Malz}, M.~{Dai}, G.~{Narayan},
  E.~E.~O. {Ishida}, Jr~{Allam}, T., A.~{Bahmanyar}, R.~{Biswas}, L.~{Galbany},
  S.~W. {Jha}, D.~O. {Jones}, R.~{Kessler}, M.~{Lochner}, A.~A. {Mahabal},
  K.~S. {Mandel}, J.~R. {Mart{\'\i}nez-Galarza}, J.~D. {McEwen},
  D.~{Muthukrishna}, H.~V. {Peiris}, C.~M. {Peters}, and C.~N. {Setzer}.
\newblock {Results of the Photometric LSST Astronomical Time-series
  Classification Challenge (PLAsTiCC)}.
\newblock \emph{arXiv e-prints}, art. arXiv:2012.12392, December 2020.

\bibitem[{Kessler} et~al.(2019){Kessler}, {Narayan}, {Avelino},
  et~al.]{Kessler19}
R.~{Kessler}, G.~{Narayan}, A.~{Avelino}, et~al.
\newblock {Models and Simulations for the Photometric LSST Astronomical Time
  Series Classification Challenge (PLAsTiCC)}.
\newblock \emph{\pasp}, 131\penalty0 (1003):\penalty0 094501, September 2019.
\newblock \doi{10.1088/1538-3873/ab26f1}.

\bibitem[{Chatterjee} et~al.(2022){Chatterjee}, {Narayan}, {Aleo}, {Malanchev},
  and {Muthukrishna}]{Chatterjee22}
Deep {Chatterjee}, Gautham {Narayan}, Patrick~D. {Aleo}, Konstantin
  {Malanchev}, and Daniel {Muthukrishna}.
\newblock {El-CID: a filter for gravitational-wave electromagnetic counterpart
  identification}.
\newblock \emph{\mnras}, 509\penalty0 (1):\penalty0 914--930, January 2022.
\newblock \doi{10.1093/mnras/stab3023}.

\bibitem[{Gagliano} et~al.(2021){Gagliano}, {Narayan}, {Engel}, {Carrasco
  Kind}, and {LSST Dark Energy Science Collaboration}]{Gagliano21}
Alex {Gagliano}, Gautham {Narayan}, Andrew {Engel}, Matias {Carrasco Kind}, and
  {LSST Dark Energy Science Collaboration}.
\newblock {GHOST: Using Only Host Galaxy Information to Accurately Associate
  and Distinguish Supernovae}.
\newblock \emph{\apj}, 908\penalty0 (2):\penalty0 170, February 2021.
\newblock \doi{10.3847/1538-4357/abd02b}.

\bibitem[{Jones} et~al.(2021){Jones}, {Foley}, {Narayan}, et~al.]{Jones21}
D.~O. {Jones}, R.~J. {Foley}, G.~{Narayan}, et~al.
\newblock {The Young Supernova Experiment: Survey Goals, Overview, and
  Operations}.
\newblock \emph{\apj}, 908\penalty0 (2):\penalty0 143, February 2021.
\newblock \doi{10.3847/1538-4357/abd7f5}.

\bibitem[{Graham} et~al.(2015){Graham}, {Foley}, {Zheng}, {Kelly}, {Shivvers},
  {Silverman}, {Filippenko}, {Clubb}, and {Ganeshalingam}]{Graham15}
M.~L. {Graham}, R.~J. {Foley}, W.~{Zheng}, P.~L. {Kelly}, I.~{Shivvers}, J.~M.
  {Silverman}, A.~V. {Filippenko}, K.~I. {Clubb}, and M.~{Ganeshalingam}.
\newblock {Twins for life? A comparative analysis of the Type Ia supernovae
  2011fe and 2011by}.
\newblock \emph{\mnras}, 446:\penalty0 2073--2088, January 2015.
\newblock \doi{10.1093/mnras/stu2221}.

\bibitem[{Stahl} et~al.(2020){Stahl}, {Mart{\'\i}nez-Palomera}, {Zheng}, {de
  Jaeger}, {Filippenko}, and {Bloom}]{Stahl:2020MNRAS.496.3553S}
Benjamin~E. {Stahl}, Jorge {Mart{\'\i}nez-Palomera}, WeiKang {Zheng}, Thomas
  {de Jaeger}, Alexei~V. {Filippenko}, and Joshua~S. {Bloom}.
\newblock {deepSIP: linking Type Ia supernova spectra to photometric quantities
  with deep learning}.
\newblock \emph{\mnras}, 496\penalty0 (3):\penalty0 3553--3571, August 2020.
\newblock \doi{10.1093/mnras/staa1706}.

\bibitem[{Chen} et~al.(2020){Chen}, {Hu}, and {Wang}]{Chen:2020ApJS..250...12C}
Xingzhuo {Chen}, Lei {Hu}, and Lifan {Wang}.
\newblock {Artificial Intelligence-Assisted Inversion (AIAI) of Synthetic Type
  Ia Supernova Spectra}.
\newblock \emph{\apjs}, 250\penalty0 (1):\penalty0 12, September 2020.
\newblock \doi{10.3847/1538-4365/ab9a3b}.

\bibitem[{Hu} et~al.(2022){Hu}, {Chen}, and {Wang}]{Hu:2022arXiv220202498H}
Lei {Hu}, Xingzhuo {Chen}, and Lifan {Wang}.
\newblock {Spectroscopic Studies of Type Ia Supernovae Using LSTM Neural
  Networks}.
\newblock \emph{arXiv e-prints}, art. arXiv:2202.02498, February 2022.

\bibitem[{Mishra} and {Molinaro}(2021)]{Mishra:2021JQSRT.27007705M}
Siddhartha {Mishra} and Roberto {Molinaro}.
\newblock {Physics informed neural networks for simulating radiative transfer}.
\newblock \emph{J. Quant. Spec. Rad. Trans.}, 270:\penalty0 107705, August
  2021.
\newblock \doi{10.1016/j.jqsrt.2021.107705}.

\bibitem[Bailey(2022)]{repository}
S.~Bailey.
\newblock Data preservation for cosmology, 2022.
\newblock Snowmass2021: Computational Frontier White Paper.

\end{thebibliography}
\end{document}